\documentclass[aps,prd,twocolumn,showpacs,eqsecnum,nofootinbib,floatfix]{revtex4-1}
\usepackage{amssymb}
\usepackage{graphicx}
\usepackage{amsmath}
\usepackage{hyperref}

\def\ts{\rm TS}

\def\aj{Astron.\ J.}

\def\mnras{Mon. Not. Roy. Astron. Soc.}

\begin{document}

\title{Bright gamma-ray Galactic Center excess and dark dwarfs: Strong
  tension for dark matter annihilation despite Milky Way halo profile}

\author{Kevork N.\ Abazajian} \email{kevork@uci.edu} 
\author{Ryan E.\ Keeley} \email{rkeeley@uci.edu} 

\affiliation{Center for
  Cosmology, Department of Physics and Astronomy, University of
  California, Irvine, Irvine, California 92697 USA}


\begin{abstract}
We incorporate Milky Way dark matter halo profile uncertainties, as
well as an accounting of diffuse gamma-ray emission uncertainties in
dark matter annihilation models for the Galactic Center Extended
gamma-ray excess (GCE) detected by the Fermi Gamma Ray Space
Telescope. The range of particle annihilation rate and masses expand
when including these unknowns. However, two of the most precise
empirical determinations of the Milky Way halo's local density and
density profile leave the signal region to be in considerable tension
with dark matter annihilation searches from combined dwarf galaxy
analyses for single-channel dark matter annihilation models.  The GCE
and dwarf tension can be alleviated if: one, the halo is very highly
concentrated or strongly contracted; two, the dark matter annihilation
signal differentiates between dwarfs and the GC; or, three, local
stellar density measures are found to be significantly lower, like
that from recent stellar counts, increasing the local dark matter
density.

\end{abstract}

\maketitle 

\begin{section}{Introduction}

The Milky Way's Galactic Center (GC) is an exceedingly crowded region
with numerous gamma-ray point sources and several sources of diffuse
emission.  It is also expected to contain a high density of dark
matter, which makes it a promising place to search for signals of dark
matter annihilation or decay.  Weakly Interacting Massive Particles
(WIMPs) are among the leading candidates for dark matter, due to a
natural mechanism for their thermal production at the proper density
in the early Universe.  Supersymmetric extensions to the standard
model of particle physics can easily accommodate a WIMP \cite{Feng:2010gw}.

In previous work, several known sources of gamma-ray emission toward
the GC have been detected and modeled.  There are 18 gamma-ray sources
within the the 7$^\circ \times$ 7$^\circ$ region about the GC within
the Second Fermi Gamma-ray LAT Source Catalog (2FGL).  For example,
the gamma-ray point source associated with Sgr A$^*$ is one of the
brightest sources in the region and its emission in this band can be
modeled as originating from hadronic cosmic rays transitioning from
diffuse to rectilinear propagation \cite{Chernyakova:2011zz}.  There
is an abundance of gamma rays associated with bremsstrahlung emission
from $e^{\pm}$, as mapped by the 20 cm radio map of the GC
\cite{YusefZadeh:2012nh}.  There is also Inverse Compton (IC) emission
that is consistent with coming from the same $e^{\pm}$ source as the
bremsstrahlung emission \cite{Abazajian:2014hsa}

After considering known sources of gamma-ray emission, there remains
an extended excess
\cite{Goodenough:2009gk,Hooper:2010mq,Hooper:2011ti,Boyarsky:2010dr,Abazajian:2012pn,Gordon:2013vta,Macias:2013vya,Calore:2014xka,Daylan:2014rsa}.
This Galactic Center Extended (GCE) excess signal gained significant
interest since it may be consistent with a WIMP dark matter
annihilation model.  Primarily, the spatial profile of the excess is
consistent with the expected profile from dark matter halos in galaxy
formation simulations.  Secondly, the strength of the signal implies
an interaction cross section that is consistent with the thermal relic
cross section.  And thirdly, the spectra of the excess signal is
consistent with a WIMP with a mass between 10-50 GeV that decays
through quark or lepton channels.  This triple consistency of the
WIMP paradigm as an explanation of the GCE has gained significant
attention.

Of course, there exist other candidates for the GCE
gamma-ray emission.  For instance, there is a large population of
compact objects which can be bright gamma-ray sources.  The GC Central
Stellar Cluster can harbor a significant population of millisecond
pulsars (MSPs).  Since MSPs can have a spectra similar to
low-particle-mass annihilating WIMPs, their presence can confuse a
dark matter interpretation of the GC emission
\cite{Baltz:2006sv,Abazajian:2010zy}. Significantly, flux probability distribution methods
have found evidence that point sources are more consistent with the
GCE flux map than a smooth halo source
\cite{Bartels:2015aea,Lee:2015fea}.

If annihilating dark matter explains the GCE, then there should be
annihilation signals in other places that have a high density of dark
matter.  Two such places are the ``inner Galaxy" (within $\sim$20$^\circ$ of
the GC) and the dwarf satellites of the Milky Way.  Previous work has
found that the inner galaxy signal is consistent with the mass and
cross section supported by the galactic center
\cite{Calore:2014xka,Daylan:2014rsa}. We will show the Milky Way dwarf
galaxies' lack of a signal
\cite{Geringer-Sameth:2014qqa,Ackermann:2015zua} significantly
constrains the GCE parameter space.  However, there is a reported
excess from the newly discovered Reticulum 2 dwarf galaxy that may be
consistent with the GC annihilation signal
\cite{Geringer-Sameth:2015lua}.  We will discuss below what would be
required to have the GCE signal be consistent with the dwarf galaxy
limits.

Previous analyses have largely used fixed values for the parameters of
the Milky Way's dark matter halo when inferring dark matter particle
properties that could produce the GCE.  There exists significant
uncertainty in these parameters, which translates into large errors on
the cross section of dark matter annihilation, while background
emission modeling uncertainties in the crowded GC region largely
generate uncertainties on the dark matter particle mass.  In this
paper, we perform a Bayesian analysis of the full GCE likelihood in
order to more properly quantify the uncertainties on the nature of
dark matter that may produce the GCE signal.  Gaussian and chi-squared
statistics are often used in other work for dark matter fits to the
GCE spectra. Such approximations are inaccurate due to the Poisson
nature of the photon count signal, and the inaccuracy is increased
when convolved with Milky Way halo uncertainties.  To assist in
particle model fits to the GCE, we also provide the tools necessary to
accurately calculate these uncertainties for general dark matter
annihilation models with arbitrary
spectra.\footnote{\href{https://github.com/rekeeley/GCE_errors}{\texttt{https://github.com/rekeeley/GCE\_errors}}}

\end{section}

\begin{figure}[t]
		\includegraphics[width=3.4truein]{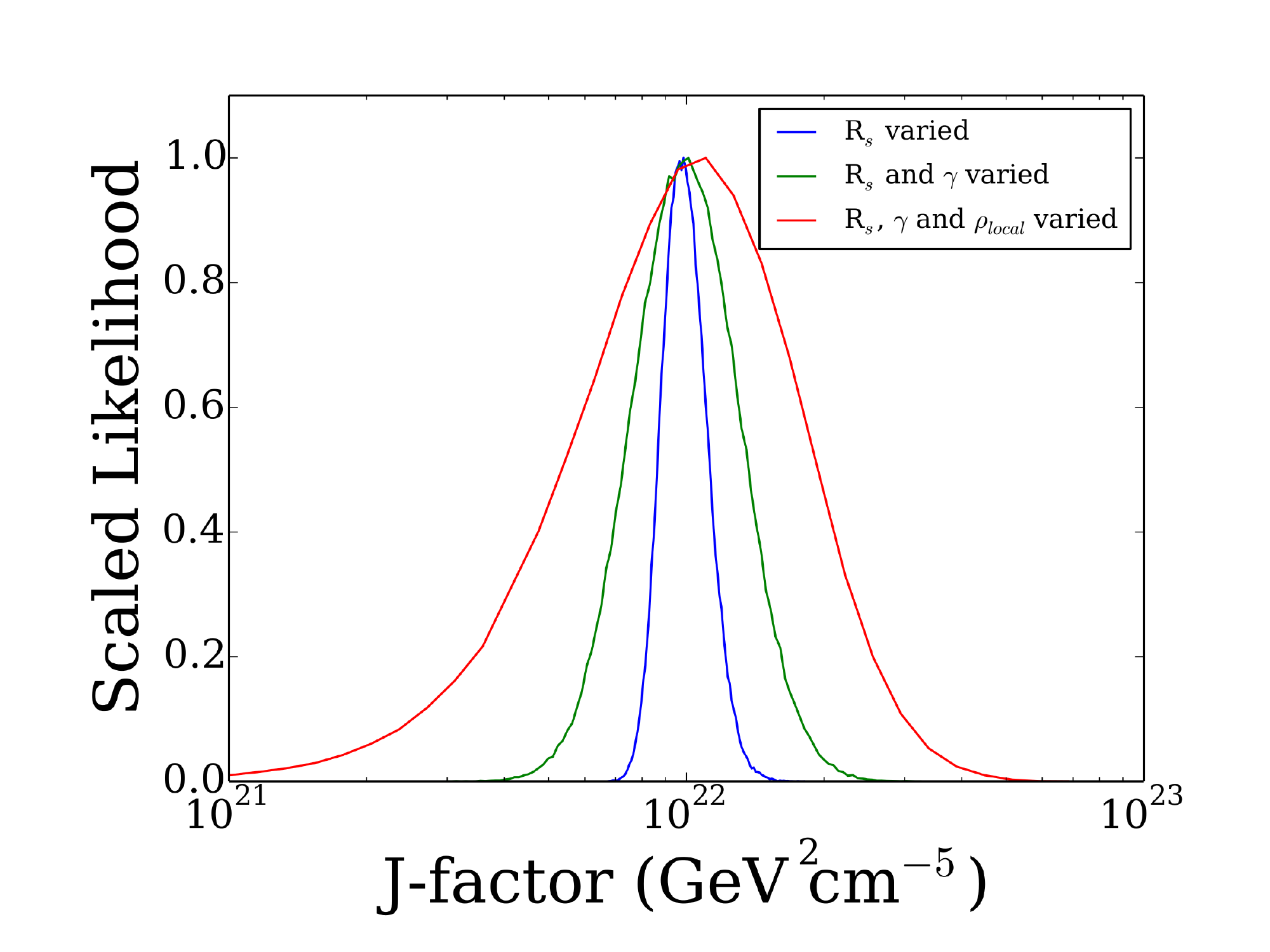}
\caption{Plotted is the scaled likelihood for the galactic center's
  J-factor for our ROI given relaxation of the constraints on the
  Milky Way dark matter halo, as described in the
  text.\label{jfactorfig}}
\end{figure}

\begin{section}{Data and Model Components}

The data set that we will refer to as the `IC' data set is taken from
the analysis in Ref.~\cite{Abazajian:2014hsa}.  It is generated with
Fermi Tools version \verb|v9r33| to study Fermi LAT observations from
August 2008 to June 2014 (approximately 70 months of data).  This data
is from Pass 7 rather than Pass 7 Reprocessed instrument response
functions since the diffuse map associated with the latter have strong
caveats for use with new extended sources.  This analysis
simultaneously fits the amplitude and spectrum of point sources from
the 2FGL catalog~\cite{Nolan2012}, plus four other point sources in
the region of interest (ROI).  It uses $0.2 - 100\ {\rm GeV}$ photons in 30
logarthmically-spaced energy bins, with \verb|ULTRACLEAN|-class photon
selection.  The IC data-set includes the 20 cm radio template as a
tracer of gas to account for the bremsstrahlung emission as has been
done previously
~\cite{YusefZadeh:2012nh,Macias:2013vya,Abazajian:2014fta}.  It also
includes IC emission from starlight with a 3.4 $\mu\rm m$ template
from the WISE mission~\cite{Wright2010}.  The IC data set also
includes the New Diffuse (ND) map whose intensity is sub-dominant to
the bremsstrahlung map and increases with angle away from the GC. The
ND template is that described in Ref.~\cite{Abazajian:2014fta}, and is
interpreted as accounting for additional bremsstrahlung emission not
captured in the 20 cm map.  The IC data set optimized the
morphology of the GCE excess and ND templates to their best-fit
profiles.  The GCE excess, used templates of density $\rho(r)^2$
projected along the line-of-sight with $\rho(r) \propto
r^{-\gamma}(r+r_s)^{-(3-\gamma)}$. The IC data analysis found that
$\gamma=1$ provided the best fit.  In this IC data set, all the 4
extended sources (GCE, ND, IC, Bremsstrahlung) were given generic
log-parabola spectral forms with four free parameters each.  The
analysis detected the WISE 3.4 $\mu\rm m$ template at very high
significance of $\ts = 197.0$\footnote{$\ts \equiv
  2\Delta\ln{\mathcal{L}}$, where $\Delta\mathcal L$ is the difference
  of the best-fit likelihood with and without the source. For point
  sources, a value of $\rm TS = 25$ is detected at a significance of
  just over $4\sigma$ \cite{Nolan2012}.}. The previously studied
sources were also detected at high significance. The GCE was detected
with $\ts = 207.5$, bremsstrahlung was detected with $\ts = 97.2$.

We adopt `noIC' and `noB' data sets from the analysis in
Ref.~\cite{Abazajian:2014fta}.  These data sets were analyzed in a
similar manner to the `IC' data, except the the `noIC' data set does
not include the inverse Compton background template, and the `noB'
includes neither the inverse Compton template nor the 20 cm radio
template.  Both these data sets cover the same $7^\circ \times
7^\circ$ ROI as the `IC' set, but use {\tt SOURCE}-class photons.
They use Fermi Tools version {\tt v9r31p1} to study Fermi LAT data
from August 2008 to May 2013 (approximately 57 months of data), and
they use Pass 7 instrument response functions.

\end{section}

\begin{section}{Analysis}

The signal strength of annihilating dark matter in the GC depends on
the density profile of the Milky Way's dark matter profile.  We the choose
dark matter density to have the generalized Navarro-Frenk-White (NFW) profile
of the form \cite{Navarro:1996gj,Klypin:2001xu}:
\begin{equation}
\rho (r) = \frac{\rho_\odot}{ \left(\frac{r} { R_\odot}\right)^\gamma
  \left(\frac{1 + r/R_s}{1 + R_\odot/R_s}\right)^{3-\gamma} },\label{nfwprofile}
\end{equation}
where $R_\odot$ is the Sun's distance from the center of the Milky
Way, $\rho_\odot$ is the density of the dark matter halo at $R_\odot$,
$R_s$ is the scale radius of the Milky Way's dark matter halo, and
$\gamma$ is a parameter characterizing the slope of the inner part of
the profile.

To arrive at substantially more accurate errors on the inferred dark
matter particle mass and cross section from the GCE signal, we employ
a Bayesian analysis to propagate uncertainties in the dark matter halo
to uncertainties in the particle annihilation parameters.  Bayesian
techniques have a formally straightforward method to include the
effect of these nuisance parameters, namely to integrate the
likelihood over the subspace of those nuisance parameters:
\begin{equation}
\mathcal{L}(\theta | x) = \int dn \ \mathcal{L} (\theta, n |x).
\end{equation} 
This defines our approach for this analysis: calculate the full
likelihood then marginalize over the nuisance subspace to get the
likelihood as a function of the dark matter mass and cross section.
The posterior distribution errors are determined and displayed via
contours of $\Delta \mathcal{L}$ that enclose the relevant credible
interval. Note that our our choice of integrand limits in
marginalization and parameter priors are always well outside the
credible intervals of the parameters of interest. The prior
distribution is either flat in logarithmic space, for the
concentration (scale radius) log-normal distribution, or flat in the
parametric value as the remaining distributions are nearly Gaussian
for the parameter values. As we shall show below, our marginalization
integral approximations leave our results to be equivalent to the
handling of nuisance parameters in frequentist statistics.

The random observable that is used in our Bayesian analysis is the
gamma-ray number counts binned by energy.  Such number counts have
Poisson statistical errors.  Hence, to do the Bayesian analysis, it is
appropriate to use a log-likelihood of the form:
\begin{equation}
\log (\mathcal{L})=\sum_i k_i \log \mu_i - \mu_i  ,
\end{equation}
up to factors that do not involve the model parameters.  Here, $k_i$
is the observed number of events in the $i$-th energy bin and $\mu_i$
is the expected number of events from the model in that energy bin.
The expected number count in bin $i$ has two components, one
associated with the dark matter annihilation, and one associated with
background sources.  The dark matter number count is given by the
integral of the spectra of the number flux over the energy bin,
multiplied by the exposure of the $i$-th bin:
\begin{equation}
\mu_i =  b_i + \epsilon_i\int_{E_i}^{E_{i+1}}\frac{d\Phi}{dE} dE ,
\end{equation}
where $b$ is the modeled background counts, $\epsilon$ is the
exposure, $d \Phi/dE$ is the differential number flux, and the
integral is over the energy bin from the observed number counts.  The
differential flux is given by:
\begin{equation}
\frac{d\Phi}{dE} =  J \frac{\langle  \sigma v \rangle}{8 \pi m_\chi^2} \frac{dN}{dE}.
\end{equation}
Here, $\langle \sigma v \rangle$ is the cross-section, $m_\chi$ is the
mass of the dark matter particle, $d N /dE$ is the per annihilation
spectra, and the J-factor is the integral of the square of the dark
matter density along the line of sight
\begin{equation}
J(\theta, \phi) = \int dz \ \rho^2(r(\theta,\phi,z)).
\end{equation}
We use the package \verb|PPPC4DMID| to generate the prompt
annihilation spectra $dN/dE$ \cite{Cirelli:2010xx}.

The largest uncertainties on dark matter particle parameters arise
from Milky Way halo parameters.  It is the Milky Way halo parameters,
$\rho_\odot, \gamma$, and $R_s$, that need to be marginalized over.
The Milky Way halo parameters are determined either from direct
observational constraints, such as that for $\rho_\odot$ and $\gamma$,
or from that expected for dark matter halos in simulations, for $R_s$,
since no significant observational constraint exists on this
scale. The dependence on $R_s$ and its uncertainty, as we shall show,
is not significant.

One robust determination of the local dark matter density is derived
from modeling the spatial and velocity distributions for a sample of
9000 K-dwarf stars from the Sloan Digital Sky Survey (SDSS) by Zhang et al.~\cite{Zhang:2012rsb}. The
velocity distribution of these stars directly measures the local
gravitational potential and, when combined with stellar density
constraints, provides a measure of the local dark matter density. The
inferred value for the local dark matter density from that work is
$\rho_\odot = 0.28 \pm 0.08\rm\ GeV\ cm^{-3}$, and we employ the exact
likelihood from that analysis. This local density is consistent
with several other determinations \cite{Read:2014qva}. 

Another recent determination of the local stellar and dark matter
density by McKee et al.~\cite{McKee:2015hwa} from star counts finds a
significantly lower total stellar mass density than the dynamical
stellar density profile measures of
Refs.~\cite{Zhang:2012rsb,Bovy:2012tw,Bovy:2013raa}. When the lower
stellar density is combined with determinations of local total mass
densities, McKee et al.\ find a higher local dark matter density
$\rho_\odot = 0.49 \pm 0.13\rm\ GeV\ cm^{-3}$. The error in McKee et
al.\ of $\sigma(\rho_\odot) = 0.13 \rm\ GeV\ cm^{-3}$ is determined
through the variation in total mass density determinations and is not
from a full error analysis. Therefore, both the error and central
value on the density from star counts are approximate.  McKee et
al.~\cite{McKee:2015hwa} also state that the dynamical estimates of
the local density like that in
Refs~\cite{Zhang:2012rsb,Bovy:2012tw,Bovy:2013raa} are the ``cleanest
determinations of the local dark matter density,'' which indicates
that perhaps the current most robust determination of the local dark
matter density to be coming from Zhang et al.~\cite{Zhang:2012rsb}.

A third recent determination of the Milky Way halo profile and local
dark matter density was done by Pato et al.\ \cite{Pato:2015dua}. Pato
et al.\ use measures of gas kinematics from neutral hydrogen terminal velocities and thickness, carbon monoxide terminal velocities, ionized hydrogen regions, and giant molecular
clouds, as well as stellar and maser kinematics. They find a larger
and more constrained value of the local density than Zhang et al., at
$\rho_\odot = 0.420^{+0.011}_{-0.009} \pm 0.025\rm\ GeV\ cm^{-3}$, while
fixing the scale radius at 20 kpc. Pato et al. find a tighter
constraint on the local dark matter density.  This results from their
constraints on models of the entirety of the Milky Way rotation curve,
with multiple kinematic sets of data to measure the local dark matter
density. Using multiple local and non-local observables to measure the
local dark matter density has the capacity to over-constrain the
dark matter profile and its local density, therefore underestimating
the true uncertainty in the local density. Exploring the multiple
constraint problem on the Milky Way's density profile from kinematic
data is beyond the scope of the work here. Therefore, we adopt the
local density in Ref.~\cite{Pato:2015dua} as a third local dark matter
density determination in our analysis. 

Other local density determinations are consistent approximately within
the range of our three density determination representative
results. For example, Refs.~\cite{Salucci:2010qr,Nesti:2013uwa} find
$\rho_\odot = 0.43^{0.11}_{-0.10}\rm\ GeV\ cm^{-3}$; while
Ref.~\cite{Iocco:2011jz} find $\rho_\odot=0.20-0.56\rm\ GeV\ cm^{-3}$ at
1$\sigma$. Our constraint from Zhang et al.~\cite{Zhang:2012rsb}
represents the lower range of density determinations, while McKee et
al.~\cite{McKee:2015hwa} represents the higher density
determinations. The framework provided here for assessing the
consistency between the GCE and dwarfs, along with the open-source
software, may be adapted to any chosen density and profile
determinations from past or future data.

The constraints on the Milky Way halo scale radius are
derived from the concentration, defined as $c \equiv R_{vir}/R_s$.
The concentration of a halo describes the scale at which the slope of
the profile of the halo changes from $\gamma$ to 3, and it has some
scatter associated with it \cite{Bullock:1999he}.  We adopt the halo
concentration's dependence on the mass of that halo as parameterized by
Sanchez-Conde \& Prada \cite{Sanchez-Conde:2013yxa}.  The
concentration is log-normally distributed with an error of 0.14 dex so
the prior likelihood for the scale radius is of the form:
\begin{equation} 
\mathcal{\log L} = -\frac{(\log_{10}(R_{vir}/R_s) - \log_{10}
  c(M_{vir}))^2 }{2 \times 0.14^2}.
\end{equation}
The concentration, which sets the scale radius, will change
with varying halo mass.  However, over a wide range of halo masses ($5
\times 10^{11} - 10^{14}$ M$_\odot$) the concentration varies only by
an amount less than the statistical variation of the concentration:
0.14 dex.  Hence, we neglect the additional uncertainty associated
with varying the halo mass.

There is some uncertainty whether the Milky Way follows a
concentration-mass relation.  Indeed, Nesti \& Salucci
\cite{Nesti:2013uwa} find that the Milky Way is an outlier and has a
value for the concentration parameter that is larger than would be
implied from Sanchez-Conde \& Prada's concentration-mass relation.
However, the scale radius found by Nesti \& Salucci is well outside
the solar radius.  In this regime, uncertainty in the solar radius
translates into a relatively small uncertainty in the J-factor.
Ultimately, the additional uncertainty introduced by Nesti \& Salucci
is bracketed by the considerations already discussed.

The inner profile of the Milky Way halo within the inner $\lesssim
500\rm\ pc$ relevant for the GCE is not well determined by dynamical
data, or numerical results, since the region becomes baryon-density
dominated. However, the profile is constrained by the observed GCE
itself. In the analysis
including bremsstrahlung emission, Abazajian et. al.
\cite{Abazajian:2014fta} find $\gamma = 1.12 \pm 0.05$. When including
the newly discovered IC component, the best-fit profile shifted to
$\gamma = 1.0$ with comparable errors \cite{Abazajian:2014hsa}.

To demonstrate the effect of allowing the parameters of the Milky
Way's dark matter halo to vary, we plot in Fig.~\ref{jfactorfig} the likelihood of the J-factor
derived from the relaxing the values of the local density, scale
radius, and slope of the inner profile.  The width of the likelihood
distribution of the J-factor expands the posterior likelihood of the
dark matter particle mass and cross section relative to using fixed
values for the halo parameters.  As Fig.~\ref{jfactorfig} shows,
varying the local density accounts for most of the width of the
J-factor likelihood, though varying the scale radius and the inner
profile slope also widens the likelihood. The J-factor likelihood is
approximately a normal distribution as it is dominated by an
approximately normal distribution in the $\rho_\odot$ uncertainty, and
sub-dominant log-normal $R_s$ and normal $\gamma$ distributions.

Because integrating the likelihoods over the nuisance subspace can be
computationally expensive, we approximate this integral by maximizing
the log-likelihood over that subspace. Since the likelihood functions
are approximately Gaussian ($\rho_\odot$ and $\gamma$) or log-normal
($R_s$) in the nuisance parameters, this is expected to be a good
approximation. We have tested that this approximation is valid by
explicitly integrating the likelihoods for single parameter
dimensions.  We explicitly calculate the probability contained within
some $\Delta \log(\mathcal{L})$ by integrating the likelihood to find
the 68\%, 95\%, and 99.7\% and 99.99997\% credible intervals for our
plotted results.  Note that the maximization of the probability
distribution leaves our results, up to an arbitrary normalization,
equivalent to the frequentist profile likelihood method of finding
statistical errors on parameters of interest when nuisance parameters
are involved.

We determine the uncertainty regions of the particle
mass and cross section parameter space for both $b$-quark and
$\tau$-lepton annihilation channels, as shown in Fig.~\ref{btaufig}.
In the next section, we investigate the systematic uncertainty
associated with uncertainties in the background-dominated low energy
data portion of the GCE, as well as uncertainties introduced by
incorporating or excluding different background diffuse emission
models, including the bremsstrahlung excess and IC component.

\end{section}

\begin{figure*}[ht!]
		\includegraphics[width=3.4truein]{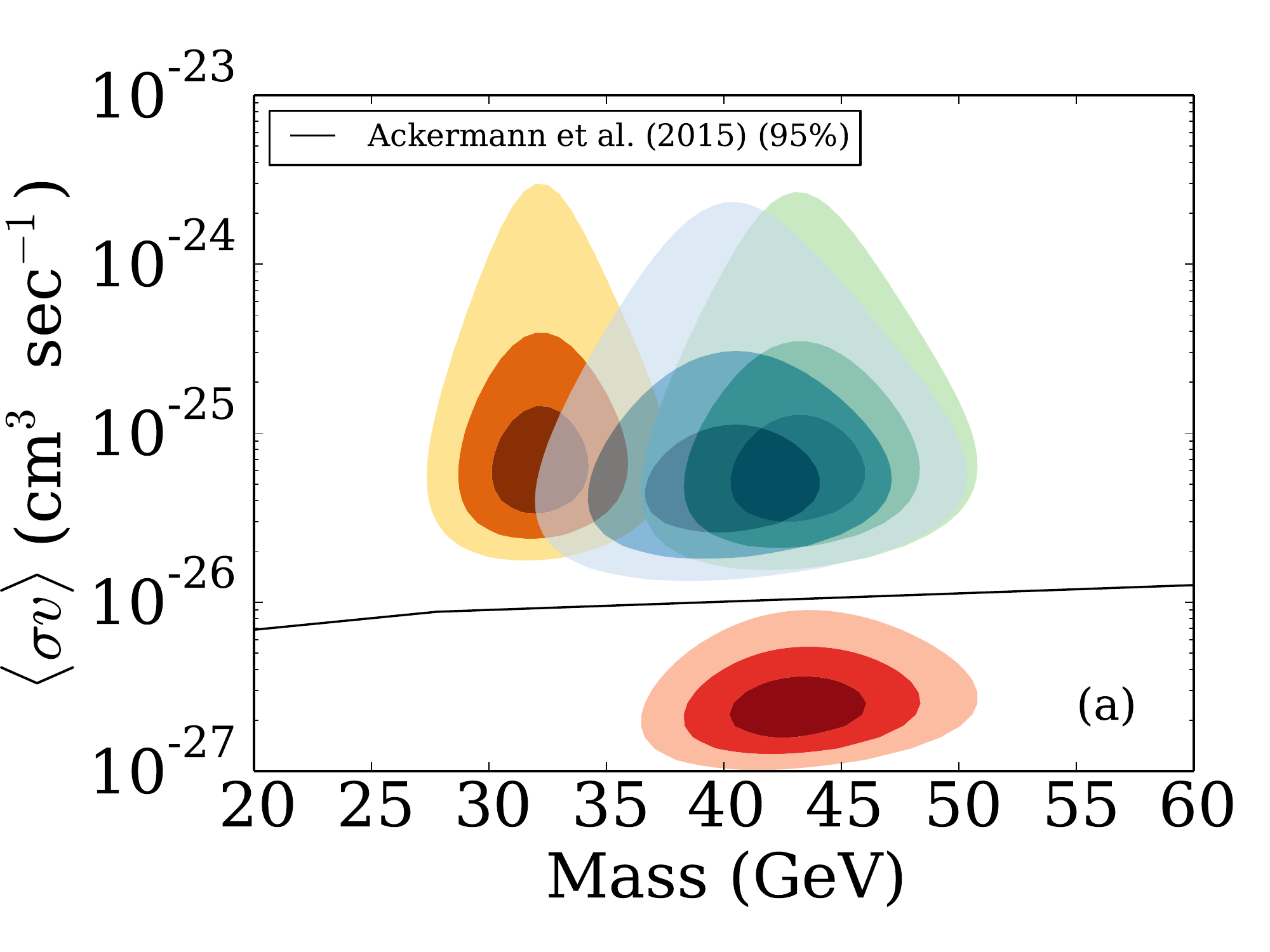}
	        ~
		\includegraphics[width=3.4truein]{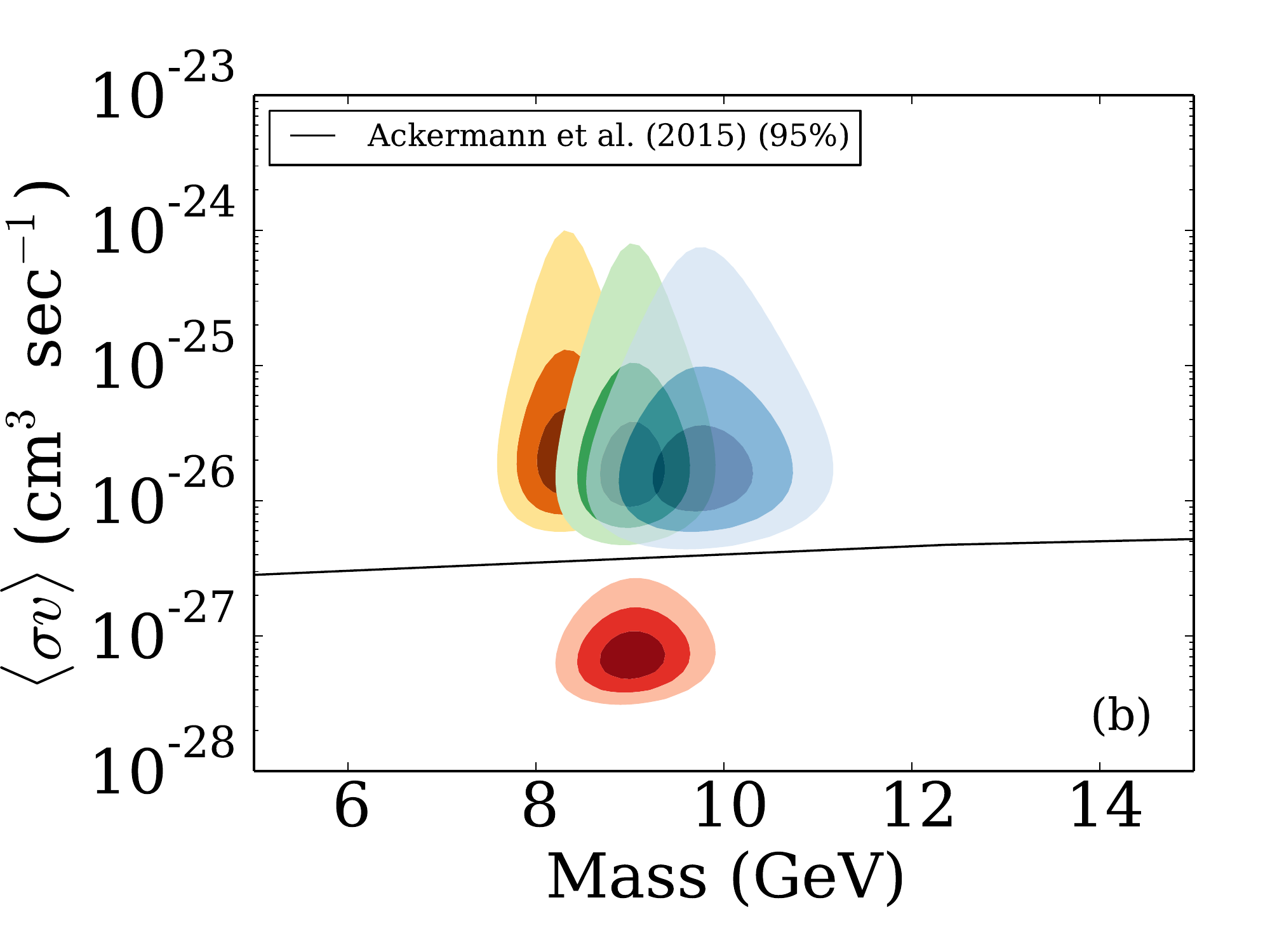}
                \\
		\includegraphics[width=3.4truein]{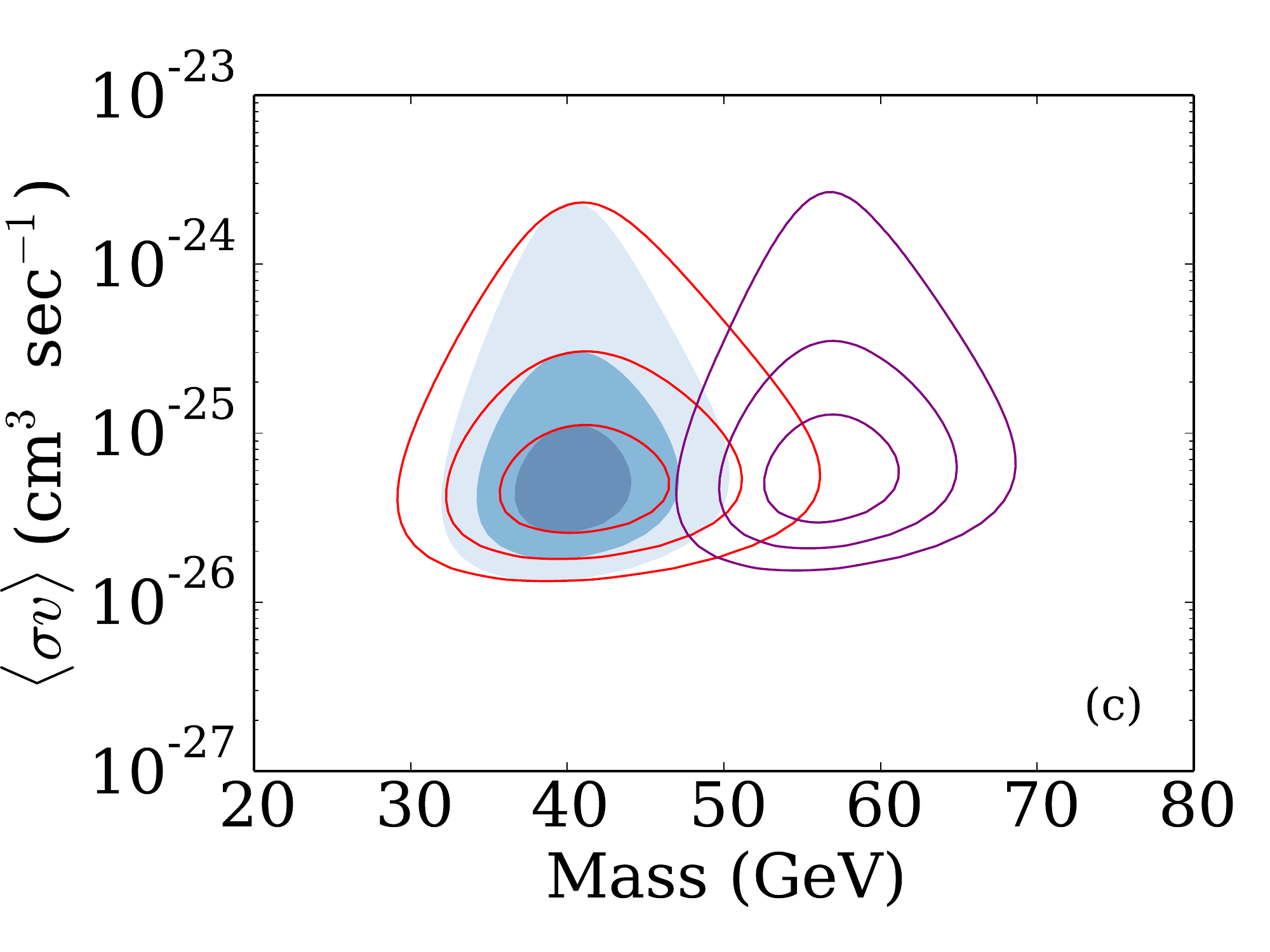}
	        ~
		\includegraphics[width=3.4truein]{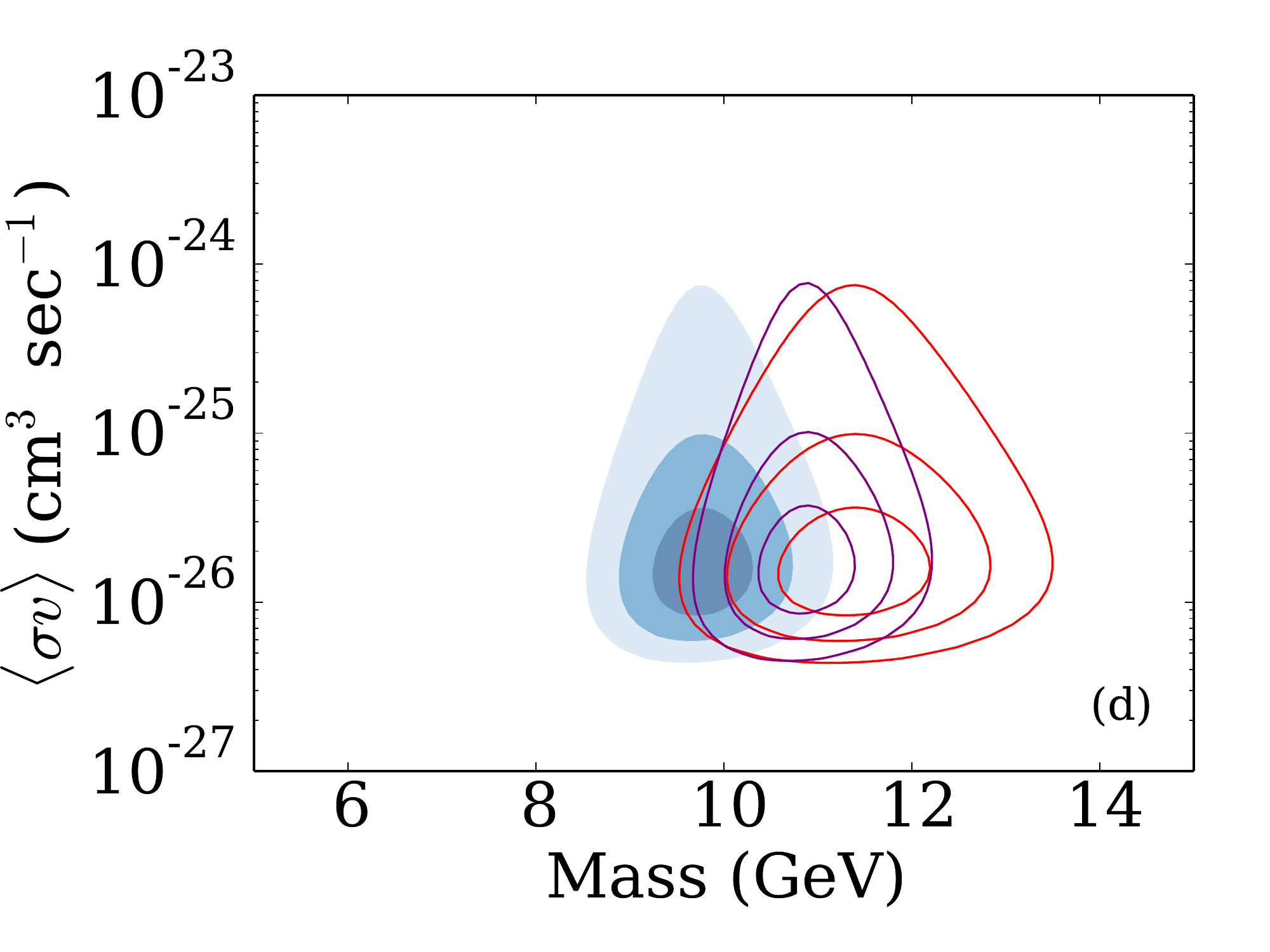}
	\caption{In (a) \& (b), we plot contours of the
          $\Delta$log-likelihood that correspond to 68\%, 95\% and
          99.7\% credible regions for the full IC, noIC, and noB
          data sets, when marginalizing over Milky Way halo
          uncertainties, which demonstrate the systematic errors
          involved in the inclusion of diffuse sources in the GC; (a)
          is for the $b/\bar b$-quark channel and (b) is for the
          $\tau^\pm$ channel. The full IC model is shown in blue, noIC
          is in orange, and noB is in green.  We also show, in red
          contours, a non-standard high-concentration/contraction Milky
          Way halo model that would escape dwarf galaxy limits, but
          would be in conflict with local density and Milky Way halo
          simulations. We also show the 95\% limits from dwarf galaxy
          searches by Ackermann et al.~\cite{Ackermann:2015zua}. In
          the (c) \& (d), for the $b/\bar b$-quark and $\tau^\pm$
          channels respectively, we plot contours of the
          $\Delta$log-likelihood that correspond to 68\%, 95\% and
          99.7\% for different numbers of low-energy bins excluded,
          demonstrating GCE spectrum determination systematic
          uncertainties in our method.  The red contours are those
          derived from excluding data below 2.03 GeV, blue from
          excluding data below 1.24 GeV, and purple with a 0.764 GeV
          cut. The blue contours are for our optimal GCE spectrum
          determination, as described in the text. \label{btaufig}}
\end{figure*}

\begin{section}{Background Diffuse Emission Model Dependence}

We test the model dependence associated with
emission from astrophysical backgrounds, including the detected
bremsstrahlung diffuse excess component and IC components producing
gamma-ray emission within the GC. Since the morphology of these
sources is not known a priori, there is a significant systematic
uncertainty introduced by the templates adopted as the model of these
diffuse sources. To bracket this model uncertainty, we take extreme
cases where the model components are either present or not. Our full
model in this work includes all components: the 20 cm bremsstrahlung,
IC, and GCE templates, as well as new diffuse and point sources as
described in Abazajian et al.~\cite{Abazajian:2014hsa}.  The noIC
(denoted `full' in Abazajian et al.~\cite{Abazajian:2014fta}) model
includes everything from the full model except the IC component.  The
noB model neglects the contribution from the 20 cm template, in
addition to neglecting the IC component.  Including different
gamma-ray source templates shifts the best-fit values of the mass,
bracketing a large part of the model dependence of the GCE emission,
as shown in the upper panels of Fig.~\ref{btaufig}. The dependence
largely in particle mass in our diffusion uncertainties and not
annihilation rate comes from the well-determined nature of the GCE
total flux at $\approx$3 GeV even for various diffuse model and GCE
spectral cases, as shown in Fig.~4 and Fig.~10 of
Ref.~\cite{Abazajian:2014fta}. Our adopted full model fit is shown in
solid colors, with the contours representing an estimate of background
uncertainties.

Additional systematic effects are associated with the low-energy data
points.  The full low-energy data in the GCE are generally not
sensitive to variations in the assumed dark matter spectra since dark
matter is sub-dominant to the background components at low energies
($< 1\rm\ GeV$); see, e.g., Fig.\ 6 of Ref.~\cite{Abazajian:2014fta}.
Since we are not performing a full template and point source fit in
this analysis, we approximate the sub-dominant nature of these
low-energy data points by excluding those that are below the flux of
other diffuse sources from our fits. In full template fits of
Refs.~\cite{Abazajian:2014fta,Abazajian:2014hsa}, the sub-dominant
flux of the GCE portion of the template at low energies does not
contribute significantly to the total fit likelihood. Including all of
these points biases the best-fit masses since the GCE errors at low
energy underestimate the full model error, and shift the best-fit dark
matter particle mass determinations relative to the full template
analysis from the same data in the full template and point source
analyses.  We investigate the bias effect by varying the the number of
low-energy data points included in the analysis. We iteratively
exclude points below 0.764 GeV, 1.24 GeV, or 2.03 GeV.  Variation of
the low-energy data point inclusion shifts the best-fit mass by
approximately 10 GeV for the $b$-quark annihilation channel, and by
around 2 GeV for the $\tau$-lepton annihilation channel, as shown in
the lower panels of Fig.~\ref{btaufig}. Including all the lower energy
data shifts to higher particle mass for the fit. Our best estimate of
the subset that represents the full template and point source analysis
is where the data simultaneously dominates above the background
sources at $\gtrsim\rm 1\ GeV$, becomes less sensitive to the number
of points included, and provides optimal sensitivity to the particle
mass, as shown in Fig.~\ref{btaufig}. The optimal case is shown in
solid colors.

Given that the parameter space for the GCE signal may be significantly
constrained by searches for annihilation in dwarf galaxies,
particularly in the Pass 8 analysis of Ref.~\cite{Ackermann:2015zua},
we explore the type of alteration of the Milky Way halo marginally
consistent with dynamical measures and allowing for a significantly
larger integrated J-factor toward the center of the galaxy: first, we
take the local density to be $\rho_\odot = 0.4\rm\ GeV\ cm^{-3}$,
which is 1.5$\sigma$ away from the constraints from Zhang et
al.~\cite{Zhang:2012rsb}; and second, we adopt the concentration to be
a highly non-standard $c=50$, which forces the scale radius of the Milky Way to
be within the $R_\odot$, boosting the inner galaxy density. Increasing
the concentration approximates a new scale possible in the dark matter
halo from baryonic effects.

NFW halos are potentially modified by the presence of baryons via
adiabatic ``contraction'' of the halos. Therefore, we also explore
this enhancement with the CONTRA tool provided by
Ref.~\cite{2011arXiv1108.5736G}. Qualitatively, the contracted profiles give
a new effective scale radius close to $R_\odot$, and a significant
enhancement of density within $R_\odot$, up to factors of
$\sim$1.5. This boosts the J-factor by $\sim$6, with a commensurate
reduction in the necessary $\langle\sigma v\rangle$ by that
amount. Therefore, the non-standard high-concentration NFW case we
propose could be plausible in some cases of contracted
profiles. Though the NFW parameters in a pure NFW sense are extreme,
the overall J-factor result is within the realm of possibility in
contracted profiles. A full scan of halo contraction involves an
analysis that exceeds the current tools like CONTRA, and is beyond the
current scope of the paper. The ``high-concentration/contraction''
case shown in Fig.~\ref{btaufig} is plausible when considering
particle physics models that directly escape the dwarf galaxy bounds.

\end{section}

\begin{figure*}[t]
		\includegraphics[width=3.4truein]{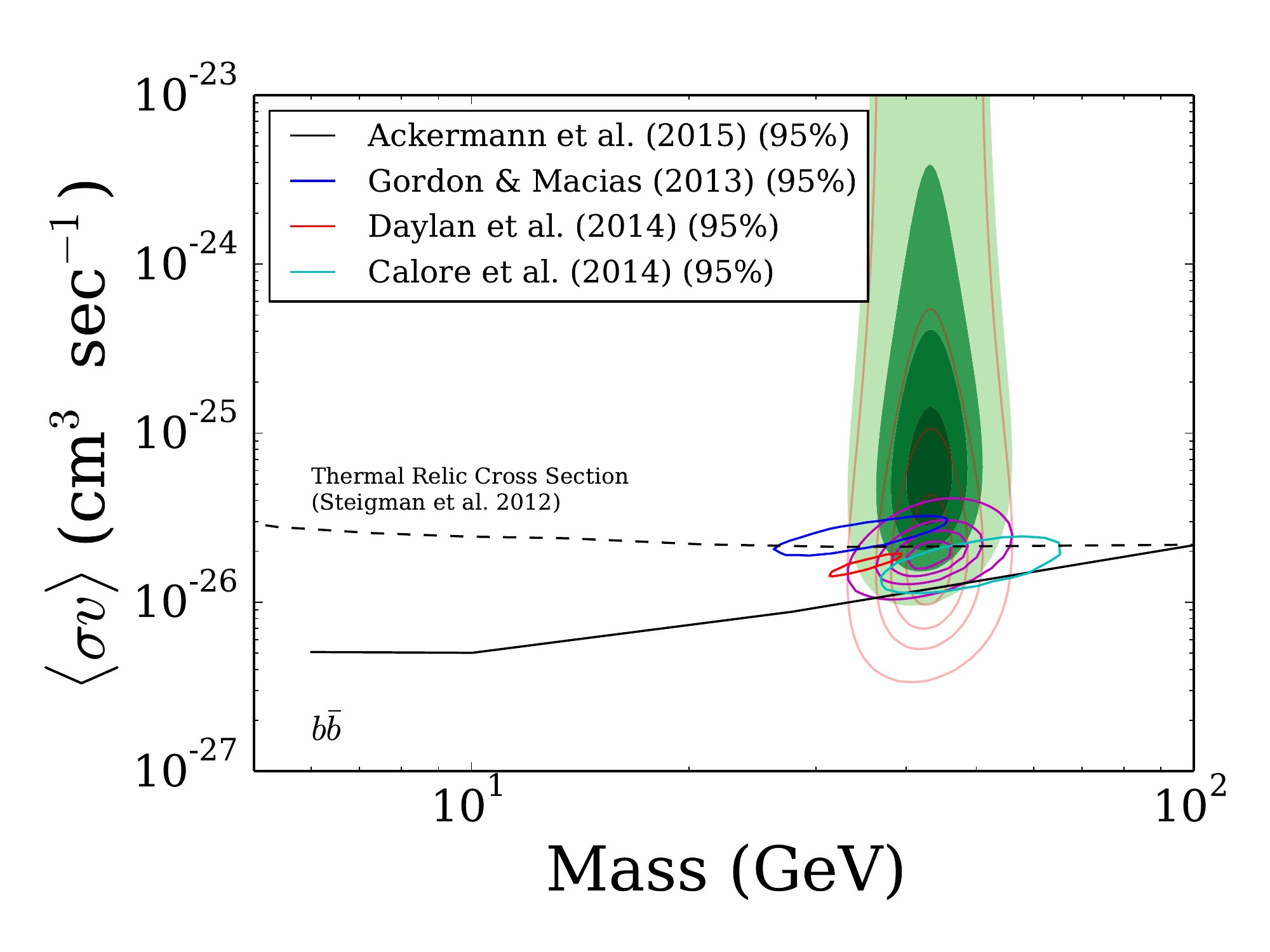}
	~
		\includegraphics[width=3.5truein]{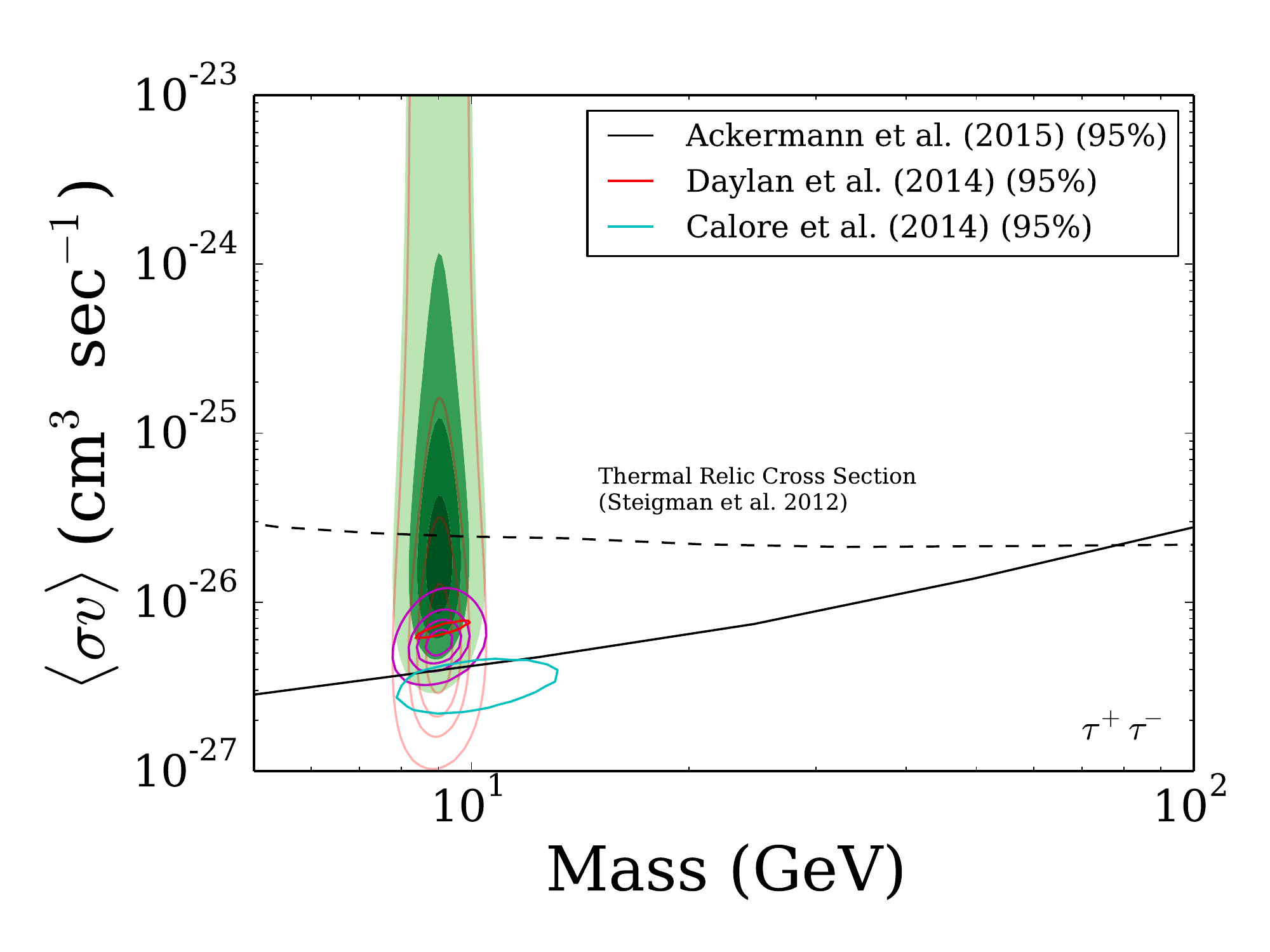}
\caption{Plotted in filled green are contours the
  $\Delta$log-likelihood that correspond to 68\%, 95\%, and 99.7\% and
  99.99997\% credible regions (corresponding approximately to 1, 2,
  3 and 5$\sigma$) when marginalizing over Milky Way halo
  uncertainties, in our best estimates for background uncertainties,
  with the local dark matter density determination by Zhang et
  al.~\cite{Zhang:2012rsb}.  Counter to the expectation that a
  symmetric error becomes asymmetric in a logarithmic plot, with
  larger extent downward, the error regions are asymmetrically
  oriented upward due to the anti-correlation of the J-factor with the
  annihilation rate $\langle \sigma v\rangle$. We also show, in light
  red, the respective approximate error contours from the inferred
  approximate dark matter density in the low stellar density star
  count measures of McKee et al.~\cite{McKee:2015hwa}.  In purple, we
  show the error contours derived from the local dark matter density
  of Pato et al.~\cite{Pato:2015dua}.  We also show the 95\% limits
  from the dwarf galaxy annihilation search by Ackermann et
  al.~\cite{Ackermann:2015zua}, and the signal regions as presented in
  Refs.~\cite{Gordon:2013vta,Daylan:2014rsa,Calore:2014xka}. As seen
  here, both the Zhang et al.\ and Pato et al.\ local dark matter
  density determinations leave single-channel dark matter annihilation
  interpretations of the GCE in strong tension with dwarf limits. The
  $b$-quark annihilation channel is on the left and the $\tau$-lepton
  annihilation channel is on the right. \label{parspacefig}}
\end{figure*}

\begin{section}{Discussion and Conclusions}
We show the credible intervals or regions consistent with three
different determinations of the local density convolved with the full
Milky Way halo profile uncertainties in Fig.~\ref{parspacefig}, and
two of the three local density determinations' parameter regions are
in significant tension with the dwarf galaxy constraints. Our results
show that allowing the local density to vary increases the errors
greatly along the cross section axis, leaving the mass axis less
constrained. This is because the effects of cross section and the
J-factor--and by extension the local density, scale radius, and inner
profile slope--are exactly inversely degenerate when fitting the data.
In particular, the inverse correlation between dark matter density and
$\langle \sigma v \rangle$ extends the error region asymmetrically
upward. This is contrast to a symmetric error in log-space, which
would extend asymmetrically downward. This illustrates the importance
of a full error analysis in quantifying uncertainties.

We also examine the background model dependence and low-energy
intensity uncertainty, which shifts the particle mass in a systematic
fashion, at the level of up to 10 GeV, depending on the overall level
of these systematic uncertainties.  We calculate the best fit dark
matter particle mass and interaction cross section implied by the GCE
that takes into account the uncertainties in the Milky Way's halo
parameters and background model uncertainties.  When adopting the SDSS
K-dwarf Zhang et al.~\cite{Zhang:2012rsb} density estimate models for
the Milky Way halo and background diffuse emission models, we found
for the $b$-quark annihilation channel that
\begin{eqnarray}
m_\chi=43. \left(^{+2.1}_{-1.9}\rm\ stat.\right)\left(\pm
19. \rm\ sys.\right)\rm\ GeV,
\\
\langle \sigma v\rangle_{bb} = 7.4
\left( ^{+2.7}_{-2.3} \right)\times 10^{-26}\rm\ cm^3\ s^{-1}.
\end{eqnarray}
For
the $\tau$-lepton channel, we found
\begin{eqnarray}
m_\chi=9.0\left(^{+0.27}_{-0.23}\rm\ stat.\right)\left(\pm
2.\rm\ sys.\right)\rm\ GeV,
\\
\langle \sigma v\rangle_{\tau} =2.2
\left( ^{+1.2}_{-0.7}\right)\times 10^{-26}\rm\ cm^3\ s^{-1}.
\end{eqnarray}
The systematic errors are defined largely by the background diffuse
emission model uncertainties, which impacts the determined dark matter
particle mass much more greatly than its cross section.  This
parameter space is significantly constrained by dwarf galaxy
annihilation searches, as shown in Fig.~\ref{parspacefig}.  The
parameter space agrees largely with other analyses.  The region found
by Calore et al.~\cite{Calore:2014xka} is a bit lower due to two
factors: they adopt a high value for $\rho_\odot =
0.4\rm\ GeV\ cm^{-3}$, as well as a more peaked central profile for
their fit at $\gamma=1.2$. A more strongly peaked central profile
$\gamma$ allows the inner and central Galaxy dark matter density to
rise to higher values, which commensurately lowers the required
annihilation rate. These modifications are along the lines of Milky
Way profile changes that would be required to escape dwarf
constraints, as discussed above.  The interaction rates for the GCE
signal at these particle masses are also being tested with collider
searches for specific couplings. For example, in the ATLAS searches for
WIMP particle production through quark couplings via vector and
axial-vector operators to dark matter constrain this region
\cite{ATLAS:2012ky}.

There are models for generation of the GCE from secondary emission of
annihilation products that could alleviate these constraints. One such
model produces the GCE as an IC emission from leptonic final states,
matching the profile and spectrum but with a significantly reduced
annihilation cross
section~\cite{Kaplinghat:2015gha,Calore:2014nla,Liu:2014cma}. The
IC-induced GCE is generated in the high value of the GC's
interstellar radiation field, while the radiation density in dwarf
galaxies is much lower, potentially allowing evasion of this tension.

Perhaps the largest systematic or modeling uncertainty is the
extrapolation of the Milky Way profile from the local density
determination, $\rho_\odot$, at $R_\odot$ to where the GCE is bright
at $\lesssim$500 pc, which is determined by the profile extrapolation
$\gamma$. For example, a strong adiabatic contraction of the Milky
Way's dark matter halo due to baryonic infall could greatly enhance
the inner Galaxy dark matter density. To illustrate a highly
non-standard, yet potentially physically viable,
high-concentration/contraction case that would be necessary to
eliminate the constraints from dwarf galaxies, we chose a high local
density and small Milky Way halo scale radius, corresponding to a high
concentration or contracted profile radius, reducing the particle dark
matter annihilation rate necessary for the GCE considerably and
avoiding the dwarf galaxy bounds. These choices for a pure NFW halo
are inconsistent with dark matter only simulations, but consistent
with halo profiles that have a contracted scale radius close to
$R_\odot$ \cite{2011arXiv1108.5736G,Abazajian:2014hsa}.  However, recent
dynamical plus microlensing data are inconsistent with a strongly
contracted halo \cite{Binney:2015gaa}. In addition, contraction is not
seen in high-mass halo systems where it is expected to more greatly
contribute~\cite{Newman:2012nv}.  Any contraction of the halo must
also preserve both the local density constraints from Zhang et
al.~\cite{Zhang:2012rsb} and the inner halo profile required by the
gamma-ray data, $\gamma = 1.0-1.2$. In summary, our
high-concentration/contraction case appears disfavored by dynamical
constraints, but evades dwarf galaxy limits and is a plausible model
for exploration of particle dark matter properties. Therefore, the
non-standard high-concentration NFW case we propose could be resolve
the tension in the case that highly contracted dark matter profiles
are found for the Milky Way.

A recent study aiming to determine the local stellar density from star
counts, McKee et al.~\cite{McKee:2015hwa}, has found lower stellar
densities than previous analyses, such as Zhang et
al.~\cite{Zhang:2012rsb}, Bovy \& Tremaine~\cite{Bovy:2012tw}, and
Bovy \& Rix \cite{Bovy:2013raa}, that determine the modeled stellar
density profile simultaneously as the dark matter profile, using the
position and velocity data of stars above the plane. If these lower
stellar densities are borne out to be accurate, with the total density
remaining invariant, then the dark matter density would be
commensurately determined to be higher. The error analysis on the
local dark matter density in McKee et al.~\cite{McKee:2015hwa} uses
the variation in total mass density determinations to set the value of
$\sigma(\rho_\odot)$ and is not the result of a full error
analysis. Therefore, both the error and central value on the density
from star counts are approximate.  

McKee et al.\ state that high-above the
Galactic plane estimates of the local density like that in
Refs~\cite{Zhang:2012rsb,Bovy:2012tw,Bovy:2013raa} are ``the cleanest
determination of the local density of dark matter,'' which indicates
the most robust determination of the local dark
matter density may be that from Zhang et
al.~\cite{Zhang:2012rsb}. However, if there is a systematic
uncertainty that shifts local stellar densities lower, our framework
and open source tools
allow for a reassessment of the GCE and dwarf agreement or tension
for arbitrary spectra of dark matter interpretations with
any new observational constraints on
Milky Way halo properties.

Another determination of the local dark matter density using a broad
set of Milky Way dynamical data was found in Pato et
al.~\cite{Pato:2015dua}. In Fig.~\ref{parspacefig} we show GCE
contours from the higher value of the approximate local dark matter
density inferred by McKee et al.~\cite{McKee:2015hwa} in light red,
and that from Pato et al.~\cite{Pato:2015dua} in purple. Importantly,
both the Zhang et al.\ and Pato et al.\ local density determinations
are inconsistent with dwarf galaxy constraints at the approximately
$\sim$5$\sigma$ level, as shown in
Fig.~\ref{parspacefig}. Significantly, it has been shown in some work
that the uncertainties in the dwarf galaxy dark matter profiles have
been underestimated, which would alleviate their constraints and
potentially relieve the GCE-dwarf tension as well
\cite{Bonnivard:2014kza}.

In summary, we performed a Bayesian analysis of the GCE emission that
more accurately accounts for uncertainties in the Milky Way halo
parameters and approximates diffuse background emission model
uncertainties.  The presence of the GCE is relatively robust to
variations in the background models, though the best fit values of the
dark matter particle mass depends significantly on these background
models.  Our analysis is certainly not an exhaustive search of all
Milky Way halo and diffuse gamma-ray emission model uncertainties, but
demonstrates the fact that uncertainties in the halo parameters
increase the uncertainty in dark matter particle
parameters. Significantly, however, we find that robust determinations
of the Milky Way halo properties, with two key determinations of the
local dark matter density \cite{Zhang:2012rsb,Pato:2015dua}, leave the
GCE parameter space in significant tension with dwarf galaxy
constraints. If the local stellar density is much higher, as in
Ref.~\cite{McKee:2015hwa}, or the Milky Way halo's dark matter density
is significantly contracted, then the tension is relaxed.  In order to
make a quantitative statement as to the level of exclusion of the GCE
by the combined dwarf analyses, a joint likelihood analysis of the
combined dwarf and GCE constraints would need to be performed.

Though the triple consistency of the dark matter interpretation of the
GCE with morphology, signal strength, and spectra remains intriguing,
the tension with dwarf galaxy annihilation searches illustrated here,
coupled with the changes to the Milky Way halo properties that would
be needed to alleviate these constraints, may indicate that
astrophysical interpretations of the GCE or more novel dark matter
annihilation mechanisms are more plausible explanations of the GCE
that are able to avoid constraints from dwarf galaxies.  Further
multiwavelength analysis is required to model background sources of
gamma-rays, which constrains the associated systematics and allows
insight into the true nature of the gamma-ray excess in the Galactic
Center.

\end{section}

\acknowledgments We thank Keith Bechtol, Mike Boylan-Kolchin, James
Bullock, Sheldon Campbell, Alex Geringer-Sameth, Manoj Kaplinghat,
Anna Kwa and Flip Tanedo for useful discussions and comments on a
draft.  K.N.A. and R.K. are partially supported by NSF CAREER Grant
No. PHY-11-59224 and NSF Grant No. PHY-1316792.

\bibliography{master}

\begin{thebibliography}{45}%
\makeatletter
\providecommand \@ifxundefined [1]{%
 \@ifx{#1\undefined}
}%
\providecommand \@ifnum [1]{%
 \ifnum #1\expandafter \@firstoftwo
 \else \expandafter \@secondoftwo
 \fi
}%
\providecommand \@ifx [1]{%
 \ifx #1\expandafter \@firstoftwo
 \else \expandafter \@secondoftwo
 \fi
}%
\providecommand \natexlab [1]{#1}%
\providecommand \enquote  [1]{``#1''}%
\providecommand \bibnamefont  [1]{#1}%
\providecommand \bibfnamefont [1]{#1}%
\providecommand \citenamefont [1]{#1}%
\providecommand \href@noop [0]{\@secondoftwo}%
\providecommand \href [0]{\begingroup \@sanitize@url \@href}%
\providecommand \@href[1]{\@@startlink{#1}\@@href}%
\providecommand \@@href[1]{\endgroup#1\@@endlink}%
\providecommand \@sanitize@url [0]{\catcode `\\12\catcode `\$12\catcode
  `\&12\catcode `\#12\catcode `\^12\catcode `\_12\catcode `\%12\relax}%
\providecommand \@@startlink[1]{}%
\providecommand \@@endlink[0]{}%
\providecommand \url  [0]{\begingroup\@sanitize@url \@url }%
\providecommand \@url [1]{\endgroup\@href {#1}{\urlprefix }}%
\providecommand \urlprefix  [0]{URL }%
\providecommand \Eprint [0]{\href }%
\providecommand \doibase [0]{http://dx.doi.org/}%
\providecommand \selectlanguage [0]{\@gobble}%
\providecommand \bibinfo  [0]{\@secondoftwo}%
\providecommand \bibfield  [0]{\@secondoftwo}%
\providecommand \translation [1]{[#1]}%
\providecommand \BibitemOpen [0]{}%
\providecommand \bibitemStop [0]{}%
\providecommand \bibitemNoStop [0]{.\EOS\space}%
\providecommand \EOS [0]{\spacefactor3000\relax}%
\providecommand \BibitemShut  [1]{\csname bibitem#1\endcsname}%
\let\auto@bib@innerbib\@empty
\bibitem [{\citenamefont {Feng}(2010)}]{Feng:2010gw}%
  \BibitemOpen
  \bibfield  {author} {\bibinfo {author} {\bibfnamefont {J.~L.}\ \bibnamefont
  {Feng}},\ }\href@noop {} {\bibfield  {journal} {\bibinfo  {journal}
  {Ann.Rev.Astron.Astrophys.}\ }\textbf {\bibinfo {volume} {48}},\ \bibinfo
  {pages} {495} (\bibinfo {year} {2010})},\ \Eprint
  {http://arxiv.org/abs/1003.0904} {arXiv:1003.0904 [astro-ph.CO]} \BibitemShut
  {NoStop}%
\bibitem [{\citenamefont {Chernyakova}\ \emph {et~al.}(2011)\citenamefont
  {Chernyakova}, \citenamefont {Malyshev}, \citenamefont {Aharonian},
  \citenamefont {Crocker},\ and\ \citenamefont {Jones}}]{Chernyakova:2011zz}%
  \BibitemOpen
  \bibfield  {author} {\bibinfo {author} {\bibfnamefont {M.}~\bibnamefont
  {Chernyakova}}, \bibinfo {author} {\bibfnamefont {D.}~\bibnamefont
  {Malyshev}}, \bibinfo {author} {\bibfnamefont {F.}~\bibnamefont {Aharonian}},
  \bibinfo {author} {\bibfnamefont {R.}~\bibnamefont {Crocker}}, \ and\
  \bibinfo {author} {\bibfnamefont {D.}~\bibnamefont {Jones}},\ }\href
  {\doibase 10.1088/0004-637X/726/2/60} {\bibfield  {journal} {\bibinfo
  {journal} {Astrophys.J.}\ }\textbf {\bibinfo {volume} {726}},\ \bibinfo
  {pages} {60} (\bibinfo {year} {2011})},\ \Eprint
  {http://arxiv.org/abs/1009.2630} {arXiv:1009.2630 [astro-ph.HE]} \BibitemShut
  {NoStop}%
\bibitem [{\citenamefont {Yusef-Zadeh}\ \emph {et~al.}(2013)\citenamefont
  {Yusef-Zadeh}, \citenamefont {Hewitt}, \citenamefont {Wardle}, \citenamefont
  {Tatischeff}, \citenamefont {Roberts} \emph {et~al.}}]{YusefZadeh:2012nh}%
  \BibitemOpen
  \bibfield  {author} {\bibinfo {author} {\bibfnamefont {F.}~\bibnamefont
  {Yusef-Zadeh}}, \bibinfo {author} {\bibfnamefont {J.}~\bibnamefont {Hewitt}},
  \bibinfo {author} {\bibfnamefont {M.}~\bibnamefont {Wardle}}, \bibinfo
  {author} {\bibfnamefont {V.}~\bibnamefont {Tatischeff}}, \bibinfo {author}
  {\bibfnamefont {D.}~\bibnamefont {Roberts}},  \emph {et~al.},\ }\href
  {\doibase 10.1088/0004-637X/762/1/33} {\bibfield  {journal} {\bibinfo
  {journal} {Astrophys.J.}\ }\textbf {\bibinfo {volume} {762}},\ \bibinfo
  {pages} {33} (\bibinfo {year} {2013})},\ \Eprint
  {http://arxiv.org/abs/1206.6882} {arXiv:1206.6882 [astro-ph.HE]} \BibitemShut
  {NoStop}%
\bibitem [{\citenamefont {Abazajian}\ \emph {et~al.}(2015)\citenamefont
  {Abazajian}, \citenamefont {Canac}, \citenamefont {Horiuchi}, \citenamefont
  {Kaplinghat},\ and\ \citenamefont {Kwa}}]{Abazajian:2014hsa}%
  \BibitemOpen
  \bibfield  {author} {\bibinfo {author} {\bibfnamefont {K.~N.}\ \bibnamefont
  {Abazajian}}, \bibinfo {author} {\bibfnamefont {N.}~\bibnamefont {Canac}},
  \bibinfo {author} {\bibfnamefont {S.}~\bibnamefont {Horiuchi}}, \bibinfo
  {author} {\bibfnamefont {M.}~\bibnamefont {Kaplinghat}}, \ and\ \bibinfo
  {author} {\bibfnamefont {A.}~\bibnamefont {Kwa}},\ }\href {\doibase
  10.1088/1475-7516/2015/07/013} {\bibfield  {journal} {\bibinfo  {journal}
  {JCAP}\ }\textbf {\bibinfo {volume} {1507}},\ \bibinfo {pages} {013}
  (\bibinfo {year} {2015})},\ \Eprint {http://arxiv.org/abs/1410.6168}
  {arXiv:1410.6168 [astro-ph.HE]} \BibitemShut {NoStop}%
\bibitem [{\citenamefont {Goodenough}\ and\ \citenamefont
  {Hooper}(2009)}]{Goodenough:2009gk}%
  \BibitemOpen
  \bibfield  {author} {\bibinfo {author} {\bibfnamefont {L.}~\bibnamefont
  {Goodenough}}\ and\ \bibinfo {author} {\bibfnamefont {D.}~\bibnamefont
  {Hooper}},\ }\href@noop {} {\  (\bibinfo {year} {2009})},\ \Eprint
  {http://arxiv.org/abs/0910.2998} {arXiv:0910.2998 [hep-ph]} \BibitemShut
  {NoStop}%
\bibitem [{\citenamefont {Hooper}\ and\ \citenamefont
  {Goodenough}(2011)}]{Hooper:2010mq}%
  \BibitemOpen
  \bibfield  {author} {\bibinfo {author} {\bibfnamefont {D.}~\bibnamefont
  {Hooper}}\ and\ \bibinfo {author} {\bibfnamefont {L.}~\bibnamefont
  {Goodenough}},\ }\href {\doibase 10.1016/j.physletb.2011.02.029} {\bibfield
  {journal} {\bibinfo  {journal} {Phys.Lett.}\ }\textbf {\bibinfo {volume}
  {B697}},\ \bibinfo {pages} {412} (\bibinfo {year} {2011})},\ \Eprint
  {http://arxiv.org/abs/1010.2752} {arXiv:1010.2752 [hep-ph]} \BibitemShut
  {NoStop}%
\bibitem [{\citenamefont {Hooper}\ and\ \citenamefont
  {Linden}(2011)}]{Hooper:2011ti}%
  \BibitemOpen
  \bibfield  {author} {\bibinfo {author} {\bibfnamefont {D.}~\bibnamefont
  {Hooper}}\ and\ \bibinfo {author} {\bibfnamefont {T.}~\bibnamefont
  {Linden}},\ }\href {\doibase 10.1103/PhysRevD.84.123005} {\bibfield
  {journal} {\bibinfo  {journal} {Phys.Rev.}\ }\textbf {\bibinfo {volume}
  {D84}},\ \bibinfo {pages} {123005} (\bibinfo {year} {2011})},\ \Eprint
  {http://arxiv.org/abs/1110.0006} {arXiv:1110.0006 [astro-ph.HE]} \BibitemShut
  {NoStop}%
\bibitem [{\citenamefont {Boyarsky}\ \emph {et~al.}(2011)\citenamefont
  {Boyarsky}, \citenamefont {Malyshev},\ and\ \citenamefont
  {Ruchayskiy}}]{Boyarsky:2010dr}%
  \BibitemOpen
  \bibfield  {author} {\bibinfo {author} {\bibfnamefont {A.}~\bibnamefont
  {Boyarsky}}, \bibinfo {author} {\bibfnamefont {D.}~\bibnamefont {Malyshev}},
  \ and\ \bibinfo {author} {\bibfnamefont {O.}~\bibnamefont {Ruchayskiy}},\
  }\href {\doibase 10.1016/j.physletb.2011.10.014} {\bibfield  {journal}
  {\bibinfo  {journal} {Phys.Lett.}\ }\textbf {\bibinfo {volume} {B705}},\
  \bibinfo {pages} {165} (\bibinfo {year} {2011})},\ \Eprint
  {http://arxiv.org/abs/1012.5839} {arXiv:1012.5839 [hep-ph]} \BibitemShut
  {NoStop}%
\bibitem [{\citenamefont {Abazajian}\ and\ \citenamefont
  {Kaplinghat}(2012)}]{Abazajian:2012pn}%
  \BibitemOpen
  \bibfield  {author} {\bibinfo {author} {\bibfnamefont {K.~N.}\ \bibnamefont
  {Abazajian}}\ and\ \bibinfo {author} {\bibfnamefont {M.}~\bibnamefont
  {Kaplinghat}},\ }\href {\doibase 10.1103/PhysRevD.86.083511} {\bibfield
  {journal} {\bibinfo  {journal} {Phys.Rev.}\ }\textbf {\bibinfo {volume}
  {D86}},\ \bibinfo {pages} {083511} (\bibinfo {year} {2012})},\ \Eprint
  {http://arxiv.org/abs/1207.6047} {arXiv:1207.6047 [astro-ph.HE]} \BibitemShut
  {NoStop}%
\bibitem [{\citenamefont {Gordon}\ and\ \citenamefont
  {Macias}(2013)}]{Gordon:2013vta}%
  \BibitemOpen
  \bibfield  {author} {\bibinfo {author} {\bibfnamefont {C.}~\bibnamefont
  {Gordon}}\ and\ \bibinfo {author} {\bibfnamefont {O.}~\bibnamefont
  {Macias}},\ }\href {\doibase 10.1103/PhysRevD.88.083521} {\bibfield
  {journal} {\bibinfo  {journal} {Phys.Rev.}\ }\textbf {\bibinfo {volume}
  {D88}},\ \bibinfo {pages} {083521} (\bibinfo {year} {2013})},\ \Eprint
  {http://arxiv.org/abs/1306.5725} {arXiv:1306.5725 [astro-ph.HE]} \BibitemShut
  {NoStop}%
\bibitem [{\citenamefont {Macias}\ and\ \citenamefont
  {Gordon}(2014)}]{Macias:2013vya}%
  \BibitemOpen
  \bibfield  {author} {\bibinfo {author} {\bibfnamefont {O.}~\bibnamefont
  {Macias}}\ and\ \bibinfo {author} {\bibfnamefont {C.}~\bibnamefont
  {Gordon}},\ }\href {\doibase 10.1103/PhysRevD.89.063515} {\bibfield
  {journal} {\bibinfo  {journal} {Phys. Rev.}\ }\textbf {\bibinfo {volume}
  {D89}},\ \bibinfo {pages} {063515} (\bibinfo {year} {2014})},\ \Eprint
  {http://arxiv.org/abs/1312.6671} {arXiv:1312.6671 [astro-ph.HE]} \BibitemShut
  {NoStop}%
\bibitem [{\citenamefont {Calore}\ \emph
  {et~al.}(2015{\natexlab{a}})\citenamefont {Calore}, \citenamefont {Cholis},\
  and\ \citenamefont {Weniger}}]{Calore:2014xka}%
  \BibitemOpen
  \bibfield  {author} {\bibinfo {author} {\bibfnamefont {F.}~\bibnamefont
  {Calore}}, \bibinfo {author} {\bibfnamefont {I.}~\bibnamefont {Cholis}}, \
  and\ \bibinfo {author} {\bibfnamefont {C.}~\bibnamefont {Weniger}},\ }\href
  {\doibase 10.1088/1475-7516/2015/03/038} {\bibfield  {journal} {\bibinfo
  {journal} {JCAP}\ }\textbf {\bibinfo {volume} {1503}},\ \bibinfo {pages}
  {038} (\bibinfo {year} {2015}{\natexlab{a}})},\ \Eprint
  {http://arxiv.org/abs/1409.0042} {arXiv:1409.0042 [astro-ph.CO]} \BibitemShut
  {NoStop}%
\bibitem [{\citenamefont {Daylan}\ \emph {et~al.}(2014)\citenamefont {Daylan},
  \citenamefont {Finkbeiner}, \citenamefont {Hooper}, \citenamefont {Linden},
  \citenamefont {Portillo} \emph {et~al.}}]{Daylan:2014rsa}%
  \BibitemOpen
  \bibfield  {author} {\bibinfo {author} {\bibfnamefont {T.}~\bibnamefont
  {Daylan}}, \bibinfo {author} {\bibfnamefont {D.~P.}\ \bibnamefont
  {Finkbeiner}}, \bibinfo {author} {\bibfnamefont {D.}~\bibnamefont {Hooper}},
  \bibinfo {author} {\bibfnamefont {T.}~\bibnamefont {Linden}}, \bibinfo
  {author} {\bibfnamefont {S.~K.~N.}\ \bibnamefont {Portillo}},  \emph
  {et~al.},\ }\href@noop {} {\  (\bibinfo {year} {2014})},\ \Eprint
  {http://arxiv.org/abs/1402.6703} {arXiv:1402.6703 [astro-ph.HE]} \BibitemShut
  {NoStop}%
\bibitem [{\citenamefont {Baltz}\ \emph {et~al.}(2007)\citenamefont {Baltz},
  \citenamefont {Taylor},\ and\ \citenamefont {Wai}}]{Baltz:2006sv}%
  \BibitemOpen
  \bibfield  {author} {\bibinfo {author} {\bibfnamefont {E.~A.}\ \bibnamefont
  {Baltz}}, \bibinfo {author} {\bibfnamefont {J.~E.}\ \bibnamefont {Taylor}}, \
  and\ \bibinfo {author} {\bibfnamefont {L.~L.}\ \bibnamefont {Wai}},\ }\href
  {\doibase 10.1086/517882} {\bibfield  {journal} {\bibinfo  {journal}
  {Astrophys. J.}\ }\textbf {\bibinfo {volume} {659}},\ \bibinfo {pages} {L125}
  (\bibinfo {year} {2007})},\ \Eprint {http://arxiv.org/abs/astro-ph/0610731}
  {arXiv:astro-ph/0610731 [astro-ph]} \BibitemShut {NoStop}%
\bibitem [{\citenamefont {Abazajian}(2011)}]{Abazajian:2010zy}%
  \BibitemOpen
  \bibfield  {author} {\bibinfo {author} {\bibfnamefont {K.~N.}\ \bibnamefont
  {Abazajian}},\ }\href {\doibase 10.1088/1475-7516/2011/03/010} {\bibfield
  {journal} {\bibinfo  {journal} {JCAP}\ }\textbf {\bibinfo {volume} {1103}},\
  \bibinfo {pages} {010} (\bibinfo {year} {2011})},\ \Eprint
  {http://arxiv.org/abs/1011.4275} {arXiv:1011.4275 [astro-ph.HE]} \BibitemShut
  {NoStop}%
\bibitem [{\citenamefont {Bartels}\ \emph {et~al.}(2015)\citenamefont
  {Bartels}, \citenamefont {Krishnamurthy},\ and\ \citenamefont
  {Weniger}}]{Bartels:2015aea}%
  \BibitemOpen
  \bibfield  {author} {\bibinfo {author} {\bibfnamefont {R.}~\bibnamefont
  {Bartels}}, \bibinfo {author} {\bibfnamefont {S.}~\bibnamefont
  {Krishnamurthy}}, \ and\ \bibinfo {author} {\bibfnamefont {C.}~\bibnamefont
  {Weniger}},\ }\href@noop {} {\  (\bibinfo {year} {2015})},\ \Eprint
  {http://arxiv.org/abs/1506.05104} {arXiv:1506.05104 [astro-ph.HE]}
  \BibitemShut {NoStop}%
\bibitem [{\citenamefont {Lee}\ \emph {et~al.}(2015)\citenamefont {Lee},
  \citenamefont {Lisanti}, \citenamefont {Safdi}, \citenamefont {Slatyer},\
  and\ \citenamefont {Xue}}]{Lee:2015fea}%
  \BibitemOpen
  \bibfield  {author} {\bibinfo {author} {\bibfnamefont {S.~K.}\ \bibnamefont
  {Lee}}, \bibinfo {author} {\bibfnamefont {M.}~\bibnamefont {Lisanti}},
  \bibinfo {author} {\bibfnamefont {B.~R.}\ \bibnamefont {Safdi}}, \bibinfo
  {author} {\bibfnamefont {T.~R.}\ \bibnamefont {Slatyer}}, \ and\ \bibinfo
  {author} {\bibfnamefont {W.}~\bibnamefont {Xue}},\ }\href@noop {} {\
  (\bibinfo {year} {2015})},\ \Eprint {http://arxiv.org/abs/1506.05124}
  {arXiv:1506.05124 [astro-ph.HE]} \BibitemShut {NoStop}%
\bibitem [{\citenamefont {Geringer-Sameth}\ \emph
  {et~al.}(2015{\natexlab{a}})\citenamefont {Geringer-Sameth}, \citenamefont
  {Koushiappas},\ and\ \citenamefont {Walker}}]{Geringer-Sameth:2014qqa}%
  \BibitemOpen
  \bibfield  {author} {\bibinfo {author} {\bibfnamefont {A.}~\bibnamefont
  {Geringer-Sameth}}, \bibinfo {author} {\bibfnamefont {S.~M.}\ \bibnamefont
  {Koushiappas}}, \ and\ \bibinfo {author} {\bibfnamefont {M.~G.}\ \bibnamefont
  {Walker}},\ }\href {\doibase 10.1103/PhysRevD.91.083535} {\bibfield
  {journal} {\bibinfo  {journal} {Phys. Rev.}\ }\textbf {\bibinfo {volume}
  {D91}},\ \bibinfo {pages} {083535} (\bibinfo {year} {2015}{\natexlab{a}})},\
  \Eprint {http://arxiv.org/abs/1410.2242} {arXiv:1410.2242 [astro-ph.CO]}
  \BibitemShut {NoStop}%
\bibitem [{\citenamefont {Ackermann}\ \emph {et~al.}(2015)\citenamefont
  {Ackermann} \emph {et~al.}}]{Ackermann:2015zua}%
  \BibitemOpen
  \bibfield  {author} {\bibinfo {author} {\bibfnamefont {M.}~\bibnamefont
  {Ackermann}} \emph {et~al.} (\bibinfo {collaboration} {Fermi-LAT}),\
  }\href@noop {} {\  (\bibinfo {year} {2015})},\ \Eprint
  {http://arxiv.org/abs/1503.02641} {arXiv:1503.02641 [astro-ph.HE]}
  \BibitemShut {NoStop}%
\bibitem [{\citenamefont {Geringer-Sameth}\ \emph
  {et~al.}(2015{\natexlab{b}})\citenamefont {Geringer-Sameth}, \citenamefont
  {Walker}, \citenamefont {Koushiappas}, \citenamefont {Koposov}, \citenamefont
  {Belokurov}, \citenamefont {Torrealba},\ and\ \citenamefont
  {Evans}}]{Geringer-Sameth:2015lua}%
  \BibitemOpen
  \bibfield  {author} {\bibinfo {author} {\bibfnamefont {A.}~\bibnamefont
  {Geringer-Sameth}}, \bibinfo {author} {\bibfnamefont {M.~G.}\ \bibnamefont
  {Walker}}, \bibinfo {author} {\bibfnamefont {S.~M.}\ \bibnamefont
  {Koushiappas}}, \bibinfo {author} {\bibfnamefont {S.~E.}\ \bibnamefont
  {Koposov}}, \bibinfo {author} {\bibfnamefont {V.}~\bibnamefont {Belokurov}},
  \bibinfo {author} {\bibfnamefont {G.}~\bibnamefont {Torrealba}}, \ and\
  \bibinfo {author} {\bibfnamefont {N.~W.}\ \bibnamefont {Evans}},\ }\href
  {\doibase 10.1103/PhysRevLett.115.081101} {\bibfield  {journal} {\bibinfo
  {journal} {Phys. Rev. Lett.}\ }\textbf {\bibinfo {volume} {115}},\ \bibinfo
  {pages} {081101} (\bibinfo {year} {2015}{\natexlab{b}})},\ \Eprint
  {http://arxiv.org/abs/1503.02320} {arXiv:1503.02320 [astro-ph.HE]}
  \BibitemShut {NoStop}%
\bibitem [{\citenamefont {{Nolan}}\ \emph {et~al.}(2012)\citenamefont
  {{Nolan}}, \citenamefont {{Abdo}}, \citenamefont {{Ackermann}}, \citenamefont
  {{Ajello}}, \citenamefont {{Allafort}}, \citenamefont {{Antolini}},
  \citenamefont {{Atwood}}, \citenamefont {{Axelsson}}, \citenamefont
  {{Baldini}}, \citenamefont {{Ballet}},\ and\ \citenamefont
  {et~al.}}]{Nolan2012}%
  \BibitemOpen
  \bibfield  {author} {\bibinfo {author} {\bibfnamefont {P.~L.}\ \bibnamefont
  {{Nolan}}}, \bibinfo {author} {\bibfnamefont {A.~A.}\ \bibnamefont {{Abdo}}},
  \bibinfo {author} {\bibfnamefont {M.}~\bibnamefont {{Ackermann}}}, \bibinfo
  {author} {\bibfnamefont {M.}~\bibnamefont {{Ajello}}}, \bibinfo {author}
  {\bibfnamefont {A.}~\bibnamefont {{Allafort}}}, \bibinfo {author}
  {\bibfnamefont {E.}~\bibnamefont {{Antolini}}}, \bibinfo {author}
  {\bibfnamefont {W.~B.}\ \bibnamefont {{Atwood}}}, \bibinfo {author}
  {\bibfnamefont {M.}~\bibnamefont {{Axelsson}}}, \bibinfo {author}
  {\bibfnamefont {L.}~\bibnamefont {{Baldini}}}, \bibinfo {author}
  {\bibfnamefont {J.}~\bibnamefont {{Ballet}}}, \ and\ \bibinfo {author}
  {\bibnamefont {et~al.}} (\bibinfo {collaboration} {Fermi-LAT
  Collaboration}),\ }\href {\doibase 10.1088/0067-0049/199/2/31} {\bibfield
  {journal} {\bibinfo  {journal} {Astrophys.J.Suppl.}\ }\textbf {\bibinfo
  {volume} {199}},\ \bibinfo {pages} {31} (\bibinfo {year} {2012})},\ \Eprint
  {http://arxiv.org/abs/1108.1435} {arXiv:1108.1435 [astro-ph.HE]} \BibitemShut
  {NoStop}%
\bibitem [{\citenamefont {Abazajian}\ \emph {et~al.}(2014)\citenamefont
  {Abazajian}, \citenamefont {Canac}, \citenamefont {Horiuchi},\ and\
  \citenamefont {Kaplinghat}}]{Abazajian:2014fta}%
  \BibitemOpen
  \bibfield  {author} {\bibinfo {author} {\bibfnamefont {K.~N.}\ \bibnamefont
  {Abazajian}}, \bibinfo {author} {\bibfnamefont {N.}~\bibnamefont {Canac}},
  \bibinfo {author} {\bibfnamefont {S.}~\bibnamefont {Horiuchi}}, \ and\
  \bibinfo {author} {\bibfnamefont {M.}~\bibnamefont {Kaplinghat}},\ }\href
  {\doibase 10.1103/PhysRevD.90.023526} {\bibfield  {journal} {\bibinfo
  {journal} {Phys.Rev.}\ }\textbf {\bibinfo {volume} {D90}},\ \bibinfo {pages}
  {023526} (\bibinfo {year} {2014})},\ \Eprint {http://arxiv.org/abs/1402.4090}
  {arXiv:1402.4090 [astro-ph.HE]} \BibitemShut {NoStop}%
\bibitem [{\citenamefont {{Wright}}\ \emph {et~al.}(2010)\citenamefont
  {{Wright}} \emph {et~al.}}]{Wright2010}%
  \BibitemOpen
  \bibfield  {author} {\bibinfo {author} {\bibfnamefont {E.~L.}\ \bibnamefont
  {{Wright}}} \emph {et~al.},\ }\href {\doibase 10.1088/0004-6256/140/6/1868}
  {\bibfield  {journal} {\bibinfo  {journal} {\aj}\ }\textbf {\bibinfo {volume}
  {140}},\ \bibinfo {eid} {1868} (\bibinfo {year} {2010})},\ \Eprint
  {http://arxiv.org/abs/1008.0031} {arXiv:1008.0031 [astro-ph.IM]} \BibitemShut
  {NoStop}%
\bibitem [{\citenamefont {Navarro}\ \emph {et~al.}(1997)\citenamefont
  {Navarro}, \citenamefont {Frenk},\ and\ \citenamefont
  {White}}]{Navarro:1996gj}%
  \BibitemOpen
  \bibfield  {author} {\bibinfo {author} {\bibfnamefont {J.~F.}\ \bibnamefont
  {Navarro}}, \bibinfo {author} {\bibfnamefont {C.~S.}\ \bibnamefont {Frenk}},
  \ and\ \bibinfo {author} {\bibfnamefont {S.~D.~M.}\ \bibnamefont {White}},\
  }\href {\doibase 10.1086/304888} {\bibfield  {journal} {\bibinfo  {journal}
  {Astrophys. J.}\ }\textbf {\bibinfo {volume} {490}},\ \bibinfo {pages} {493}
  (\bibinfo {year} {1997})},\ \Eprint {http://arxiv.org/abs/astro-ph/9611107}
  {arXiv:astro-ph/9611107} \BibitemShut {NoStop}%
\bibitem [{\citenamefont {Klypin}\ \emph {et~al.}(2002)\citenamefont {Klypin},
  \citenamefont {Zhao},\ and\ \citenamefont {Somerville}}]{Klypin:2001xu}%
  \BibitemOpen
  \bibfield  {author} {\bibinfo {author} {\bibfnamefont {A.}~\bibnamefont
  {Klypin}}, \bibinfo {author} {\bibfnamefont {H.}~\bibnamefont {Zhao}}, \ and\
  \bibinfo {author} {\bibfnamefont {R.~S.}\ \bibnamefont {Somerville}},\ }\href
  {\doibase 10.1086/340656} {\bibfield  {journal} {\bibinfo  {journal}
  {Astrophys. J.}\ }\textbf {\bibinfo {volume} {573}},\ \bibinfo {pages} {597}
  (\bibinfo {year} {2002})},\ \Eprint {http://arxiv.org/abs/astro-ph/0110390}
  {arXiv:astro-ph/0110390} \BibitemShut {NoStop}%
\bibitem [{\citenamefont {Cirelli}\ \emph {et~al.}(2011)\citenamefont
  {Cirelli}, \citenamefont {Corcella}, \citenamefont {Hektor}, \citenamefont
  {Hutsi}, \citenamefont {Kadastik} \emph {et~al.}}]{Cirelli:2010xx}%
  \BibitemOpen
  \bibfield  {author} {\bibinfo {author} {\bibfnamefont {M.}~\bibnamefont
  {Cirelli}}, \bibinfo {author} {\bibfnamefont {G.}~\bibnamefont {Corcella}},
  \bibinfo {author} {\bibfnamefont {A.}~\bibnamefont {Hektor}}, \bibinfo
  {author} {\bibfnamefont {G.}~\bibnamefont {Hutsi}}, \bibinfo {author}
  {\bibfnamefont {M.}~\bibnamefont {Kadastik}},  \emph {et~al.},\ }\href
  {\doibase 10.1088/1475-7516/2012/10/E01, 10.1088/1475-7516/2011/03/051}
  {\bibfield  {journal} {\bibinfo  {journal} {JCAP}\ }\textbf {\bibinfo
  {volume} {1103}},\ \bibinfo {pages} {051} (\bibinfo {year} {2011})},\ \Eprint
  {http://arxiv.org/abs/1012.4515} {arXiv:1012.4515 [hep-ph]} \BibitemShut
  {NoStop}%
\bibitem [{\citenamefont {Zhang}\ \emph {et~al.}(2013)\citenamefont {Zhang},
  \citenamefont {Rix}, \citenamefont {van~de Ven}, \citenamefont {Bovy},
  \citenamefont {Liu},\ and\ \citenamefont {Zhao}}]{Zhang:2012rsb}%
  \BibitemOpen
  \bibfield  {author} {\bibinfo {author} {\bibfnamefont {L.}~\bibnamefont
  {Zhang}}, \bibinfo {author} {\bibfnamefont {H.-W.}\ \bibnamefont {Rix}},
  \bibinfo {author} {\bibfnamefont {G.}~\bibnamefont {van~de Ven}}, \bibinfo
  {author} {\bibfnamefont {J.}~\bibnamefont {Bovy}}, \bibinfo {author}
  {\bibfnamefont {C.}~\bibnamefont {Liu}}, \ and\ \bibinfo {author}
  {\bibfnamefont {G.}~\bibnamefont {Zhao}},\ }\href {\doibase
  10.1088/0004-637X/772/2/108} {\bibfield  {journal} {\bibinfo  {journal}
  {Astrophys. J.}\ }\textbf {\bibinfo {volume} {772}},\ \bibinfo {pages} {108}
  (\bibinfo {year} {2013})},\ \Eprint {http://arxiv.org/abs/1209.0256}
  {arXiv:1209.0256 [astro-ph.GA]} \BibitemShut {NoStop}%
\bibitem [{\citenamefont {Read}(2014)}]{Read:2014qva}%
  \BibitemOpen
  \bibfield  {author} {\bibinfo {author} {\bibfnamefont {J.~I.}\ \bibnamefont
  {Read}},\ }\href {\doibase 10.1088/0954-3899/41/6/063101} {\bibfield
  {journal} {\bibinfo  {journal} {J. Phys.}\ }\textbf {\bibinfo {volume}
  {G41}},\ \bibinfo {pages} {063101} (\bibinfo {year} {2014})},\ \Eprint
  {http://arxiv.org/abs/1404.1938} {arXiv:1404.1938 [astro-ph.GA]} \BibitemShut
  {NoStop}%
\bibitem [{\citenamefont {McKee}\ \emph {et~al.}(2015)\citenamefont {McKee},
  \citenamefont {Parravano},\ and\ \citenamefont {Hollenbach}}]{McKee:2015hwa}%
  \BibitemOpen
  \bibfield  {author} {\bibinfo {author} {\bibfnamefont {C.~F.}\ \bibnamefont
  {McKee}}, \bibinfo {author} {\bibfnamefont {A.}~\bibnamefont {Parravano}}, \
  and\ \bibinfo {author} {\bibfnamefont {D.~J.}\ \bibnamefont {Hollenbach}},\
  }\href {\doibase 10.1088/0004-637X, 10.1088/0004-637X/814/1/13} {\bibfield
  {journal} {\bibinfo  {journal} {Astrophys. J.}\ }\textbf {\bibinfo {volume}
  {814}},\ \bibinfo {pages} {13} (\bibinfo {year} {2015})},\ \Eprint
  {http://arxiv.org/abs/1509.05334} {arXiv:1509.05334 [astro-ph.GA]}
  \BibitemShut {NoStop}%
\bibitem [{\citenamefont {Bovy}\ and\ \citenamefont
  {Tremaine}(2012)}]{Bovy:2012tw}%
  \BibitemOpen
  \bibfield  {author} {\bibinfo {author} {\bibfnamefont {J.}~\bibnamefont
  {Bovy}}\ and\ \bibinfo {author} {\bibfnamefont {S.}~\bibnamefont
  {Tremaine}},\ }\href {\doibase 10.1088/0004-637X/756/1/89} {\bibfield
  {journal} {\bibinfo  {journal} {Astrophys.J.}\ }\textbf {\bibinfo {volume}
  {756}},\ \bibinfo {pages} {89} (\bibinfo {year} {2012})},\ \Eprint
  {http://arxiv.org/abs/1205.4033} {arXiv:1205.4033 [astro-ph.GA]} \BibitemShut
  {NoStop}%
\bibitem [{\citenamefont {Bovy}\ and\ \citenamefont
  {Rix}(2013)}]{Bovy:2013raa}%
  \BibitemOpen
  \bibfield  {author} {\bibinfo {author} {\bibfnamefont {J.}~\bibnamefont
  {Bovy}}\ and\ \bibinfo {author} {\bibfnamefont {H.-W.}\ \bibnamefont {Rix}},\
  }\href {\doibase 10.1088/0004-637X/779/2/115} {\bibfield  {journal} {\bibinfo
   {journal} {Astrophys. J.}\ }\textbf {\bibinfo {volume} {779}},\ \bibinfo
  {pages} {115} (\bibinfo {year} {2013})},\ \Eprint
  {http://arxiv.org/abs/1309.0809} {arXiv:1309.0809 [astro-ph.GA]} \BibitemShut
  {NoStop}%
\bibitem [{\citenamefont {Pato}\ \emph {et~al.}(2015)\citenamefont {Pato},
  \citenamefont {Iocco},\ and\ \citenamefont {Bertone}}]{Pato:2015dua}%
  \BibitemOpen
  \bibfield  {author} {\bibinfo {author} {\bibfnamefont {M.}~\bibnamefont
  {Pato}}, \bibinfo {author} {\bibfnamefont {F.}~\bibnamefont {Iocco}}, \ and\
  \bibinfo {author} {\bibfnamefont {G.}~\bibnamefont {Bertone}},\ }\href
  {\doibase 10.1088/1475-7516/2015/12/001} {\bibfield  {journal} {\bibinfo
  {journal} {JCAP}\ }\textbf {\bibinfo {volume} {1512}},\ \bibinfo {pages}
  {001} (\bibinfo {year} {2015})},\ \Eprint {http://arxiv.org/abs/1504.06324}
  {arXiv:1504.06324 [astro-ph.GA]} \BibitemShut {NoStop}%
\bibitem [{\citenamefont {Salucci}\ \emph {et~al.}(2010)\citenamefont
  {Salucci}, \citenamefont {Nesti}, \citenamefont {Gentile},\ and\
  \citenamefont {Martins}}]{Salucci:2010qr}%
  \BibitemOpen
  \bibfield  {author} {\bibinfo {author} {\bibfnamefont {P.}~\bibnamefont
  {Salucci}}, \bibinfo {author} {\bibfnamefont {F.}~\bibnamefont {Nesti}},
  \bibinfo {author} {\bibfnamefont {G.}~\bibnamefont {Gentile}}, \ and\
  \bibinfo {author} {\bibfnamefont {C.~F.}\ \bibnamefont {Martins}},\ }\href
  {\doibase 10.1051/0004-6361/201014385} {\bibfield  {journal} {\bibinfo
  {journal} {Astron. Astrophys.}\ }\textbf {\bibinfo {volume} {523}},\ \bibinfo
  {pages} {A83} (\bibinfo {year} {2010})},\ \Eprint
  {http://arxiv.org/abs/1003.3101} {arXiv:1003.3101 [astro-ph.GA]} \BibitemShut
  {NoStop}%
\bibitem [{\citenamefont {Nesti}\ and\ \citenamefont
  {Salucci}(2013)}]{Nesti:2013uwa}%
  \BibitemOpen
  \bibfield  {author} {\bibinfo {author} {\bibfnamefont {F.}~\bibnamefont
  {Nesti}}\ and\ \bibinfo {author} {\bibfnamefont {P.}~\bibnamefont
  {Salucci}},\ }\href {\doibase 10.1088/1475-7516/2013/07/016} {\bibfield
  {journal} {\bibinfo  {journal} {JCAP}\ }\textbf {\bibinfo {volume} {1307}},\
  \bibinfo {pages} {016} (\bibinfo {year} {2013})},\ \Eprint
  {http://arxiv.org/abs/1304.5127} {arXiv:1304.5127 [astro-ph.GA]} \BibitemShut
  {NoStop}%
\bibitem [{\citenamefont {Iocco}\ \emph {et~al.}(2011)\citenamefont {Iocco},
  \citenamefont {Pato}, \citenamefont {Bertone},\ and\ \citenamefont
  {Jetzer}}]{Iocco:2011jz}%
  \BibitemOpen
  \bibfield  {author} {\bibinfo {author} {\bibfnamefont {F.}~\bibnamefont
  {Iocco}}, \bibinfo {author} {\bibfnamefont {M.}~\bibnamefont {Pato}},
  \bibinfo {author} {\bibfnamefont {G.}~\bibnamefont {Bertone}}, \ and\
  \bibinfo {author} {\bibfnamefont {P.}~\bibnamefont {Jetzer}},\ }\href
  {\doibase 10.1088/1475-7516/2011/11/029} {\bibfield  {journal} {\bibinfo
  {journal} {JCAP}\ }\textbf {\bibinfo {volume} {1111}},\ \bibinfo {pages}
  {029} (\bibinfo {year} {2011})},\ \Eprint {http://arxiv.org/abs/1107.5810}
  {arXiv:1107.5810 [astro-ph.GA]} \BibitemShut {NoStop}%
\bibitem [{\citenamefont {Bullock}\ \emph {et~al.}(2001)\citenamefont
  {Bullock}, \citenamefont {Kolatt}, \citenamefont {Sigad}, \citenamefont
  {Somerville}, \citenamefont {Kravtsov}, \citenamefont {Klypin}, \citenamefont
  {Primack},\ and\ \citenamefont {Dekel}}]{Bullock:1999he}%
  \BibitemOpen
  \bibfield  {author} {\bibinfo {author} {\bibfnamefont {J.~S.}\ \bibnamefont
  {Bullock}}, \bibinfo {author} {\bibfnamefont {T.~S.}\ \bibnamefont {Kolatt}},
  \bibinfo {author} {\bibfnamefont {Y.}~\bibnamefont {Sigad}}, \bibinfo
  {author} {\bibfnamefont {R.~S.}\ \bibnamefont {Somerville}}, \bibinfo
  {author} {\bibfnamefont {A.~V.}\ \bibnamefont {Kravtsov}}, \bibinfo {author}
  {\bibfnamefont {A.~A.}\ \bibnamefont {Klypin}}, \bibinfo {author}
  {\bibfnamefont {J.~R.}\ \bibnamefont {Primack}}, \ and\ \bibinfo {author}
  {\bibfnamefont {A.}~\bibnamefont {Dekel}},\ }\href {\doibase
  10.1046/j.1365-8711.2001.04068.x} {\bibfield  {journal} {\bibinfo  {journal}
  {Mon. Not. Roy. Astron. Soc.}\ }\textbf {\bibinfo {volume} {321}},\ \bibinfo
  {pages} {559} (\bibinfo {year} {2001})},\ \Eprint
  {http://arxiv.org/abs/astro-ph/9908159} {arXiv:astro-ph/9908159 [astro-ph]}
  \BibitemShut {NoStop}%
\bibitem [{\citenamefont {Sánchez-Conde}\ and\ \citenamefont
  {Prada}(2014)}]{Sanchez-Conde:2013yxa}%
  \BibitemOpen
  \bibfield  {author} {\bibinfo {author} {\bibfnamefont {M.~A.}\ \bibnamefont
  {Sánchez-Conde}}\ and\ \bibinfo {author} {\bibfnamefont {F.}~\bibnamefont
  {Prada}},\ }\href {\doibase 10.1093/mnras/stu1014} {\bibfield  {journal}
  {\bibinfo  {journal} {Mon. Not. Roy. Astron. Soc.}\ }\textbf {\bibinfo
  {volume} {442}},\ \bibinfo {pages} {2271} (\bibinfo {year} {2014})},\ \Eprint
  {http://arxiv.org/abs/1312.1729} {arXiv:1312.1729 [astro-ph.CO]} \BibitemShut
  {NoStop}%
\bibitem [{\citenamefont {{Gnedin}}\ \emph {et~al.}(2011)\citenamefont
  {{Gnedin}}, \citenamefont {{Ceverino}}, \citenamefont {{Gnedin}},
  \citenamefont {{Klypin}}, \citenamefont {{Kravtsov}}, \citenamefont
  {{Levine}}, \citenamefont {{Nagai}},\ and\ \citenamefont
  {{Yepes}}}]{2011arXiv1108.5736G}%
  \BibitemOpen
  \bibfield  {author} {\bibinfo {author} {\bibfnamefont {O.~Y.}\ \bibnamefont
  {{Gnedin}}}, \bibinfo {author} {\bibfnamefont {D.}~\bibnamefont
  {{Ceverino}}}, \bibinfo {author} {\bibfnamefont {N.~Y.}\ \bibnamefont
  {{Gnedin}}}, \bibinfo {author} {\bibfnamefont {A.~A.}\ \bibnamefont
  {{Klypin}}}, \bibinfo {author} {\bibfnamefont {A.~V.}\ \bibnamefont
  {{Kravtsov}}}, \bibinfo {author} {\bibfnamefont {R.}~\bibnamefont
  {{Levine}}}, \bibinfo {author} {\bibfnamefont {D.}~\bibnamefont {{Nagai}}}, \
  and\ \bibinfo {author} {\bibfnamefont {G.}~\bibnamefont {{Yepes}}},\
  }\href@noop {} {\bibfield  {journal} {\bibinfo  {journal} {ArXiv e-prints}\ }
  (\bibinfo {year} {2011})},\ \Eprint {http://arxiv.org/abs/1108.5736}
  {arXiv:1108.5736 [astro-ph.CO]} \BibitemShut {NoStop}%
\bibitem [{\citenamefont {Aad}\ \emph {et~al.}(2013)\citenamefont {Aad} \emph
  {et~al.}}]{ATLAS:2012ky}%
  \BibitemOpen
  \bibfield  {author} {\bibinfo {author} {\bibfnamefont {G.}~\bibnamefont
  {Aad}} \emph {et~al.} (\bibinfo {collaboration} {ATLAS}),\ }\href {\doibase
  10.1007/JHEP04(2013)075} {\bibfield  {journal} {\bibinfo  {journal} {JHEP}\
  }\textbf {\bibinfo {volume} {04}},\ \bibinfo {pages} {075} (\bibinfo {year}
  {2013})},\ \Eprint {http://arxiv.org/abs/1210.4491} {arXiv:1210.4491
  [hep-ex]} \BibitemShut {NoStop}%
\bibitem [{\citenamefont {Kaplinghat}\ \emph {et~al.}(2015)\citenamefont
  {Kaplinghat}, \citenamefont {Linden},\ and\ \citenamefont
  {Yu}}]{Kaplinghat:2015gha}%
  \BibitemOpen
  \bibfield  {author} {\bibinfo {author} {\bibfnamefont {M.}~\bibnamefont
  {Kaplinghat}}, \bibinfo {author} {\bibfnamefont {T.}~\bibnamefont {Linden}},
  \ and\ \bibinfo {author} {\bibfnamefont {H.-B.}\ \bibnamefont {Yu}},\ }\href
  {\doibase 10.1103/PhysRevLett.114.211303} {\bibfield  {journal} {\bibinfo
  {journal} {Phys. Rev. Lett.}\ }\textbf {\bibinfo {volume} {114}},\ \bibinfo
  {pages} {211303} (\bibinfo {year} {2015})},\ \Eprint
  {http://arxiv.org/abs/1501.03507} {arXiv:1501.03507 [hep-ph]} \BibitemShut
  {NoStop}%
\bibitem [{\citenamefont {Calore}\ \emph
  {et~al.}(2015{\natexlab{b}})\citenamefont {Calore}, \citenamefont {Cholis},
  \citenamefont {McCabe},\ and\ \citenamefont {Weniger}}]{Calore:2014nla}%
  \BibitemOpen
  \bibfield  {author} {\bibinfo {author} {\bibfnamefont {F.}~\bibnamefont
  {Calore}}, \bibinfo {author} {\bibfnamefont {I.}~\bibnamefont {Cholis}},
  \bibinfo {author} {\bibfnamefont {C.}~\bibnamefont {McCabe}}, \ and\ \bibinfo
  {author} {\bibfnamefont {C.}~\bibnamefont {Weniger}},\ }\href {\doibase
  10.1103/PhysRevD.91.063003} {\bibfield  {journal} {\bibinfo  {journal} {Phys.
  Rev.}\ }\textbf {\bibinfo {volume} {D91}},\ \bibinfo {pages} {063003}
  (\bibinfo {year} {2015}{\natexlab{b}})},\ \Eprint
  {http://arxiv.org/abs/1411.4647} {arXiv:1411.4647 [hep-ph]} \BibitemShut
  {NoStop}%
\bibitem [{\citenamefont {Liu}\ \emph {et~al.}(2015)\citenamefont {Liu},
  \citenamefont {Weiner},\ and\ \citenamefont {Xue}}]{Liu:2014cma}%
  \BibitemOpen
  \bibfield  {author} {\bibinfo {author} {\bibfnamefont {J.}~\bibnamefont
  {Liu}}, \bibinfo {author} {\bibfnamefont {N.}~\bibnamefont {Weiner}}, \ and\
  \bibinfo {author} {\bibfnamefont {W.}~\bibnamefont {Xue}},\ }\href {\doibase
  10.1007/JHEP08(2015)050} {\bibfield  {journal} {\bibinfo  {journal} {JHEP}\
  }\textbf {\bibinfo {volume} {08}},\ \bibinfo {pages} {050} (\bibinfo {year}
  {2015})},\ \Eprint {http://arxiv.org/abs/1412.1485} {arXiv:1412.1485
  [hep-ph]} \BibitemShut {NoStop}%
\bibitem [{\citenamefont {Binney}\ and\ \citenamefont
  {Piffl}(2015)}]{Binney:2015gaa}%
  \BibitemOpen
  \bibfield  {author} {\bibinfo {author} {\bibfnamefont {J.}~\bibnamefont
  {Binney}}\ and\ \bibinfo {author} {\bibfnamefont {T.}~\bibnamefont {Piffl}},\
  }\href@noop {} {\  (\bibinfo {year} {2015})},\ \bibinfo {note} {\mnras
  accepted},\ \Eprint {http://arxiv.org/abs/1509.06877} {arXiv:1509.06877
  [astro-ph.GA]} \BibitemShut {NoStop}%
\bibitem [{\citenamefont {Newman}\ \emph {et~al.}(2013)\citenamefont {Newman},
  \citenamefont {Treu}, \citenamefont {Ellis}, \citenamefont {Sand},
  \citenamefont {Nipoti}, \citenamefont {Richard},\ and\ \citenamefont
  {Jullo}}]{Newman:2012nv}%
  \BibitemOpen
  \bibfield  {author} {\bibinfo {author} {\bibfnamefont {A.~B.}\ \bibnamefont
  {Newman}}, \bibinfo {author} {\bibfnamefont {T.}~\bibnamefont {Treu}},
  \bibinfo {author} {\bibfnamefont {R.~S.}\ \bibnamefont {Ellis}}, \bibinfo
  {author} {\bibfnamefont {D.~J.}\ \bibnamefont {Sand}}, \bibinfo {author}
  {\bibfnamefont {C.}~\bibnamefont {Nipoti}}, \bibinfo {author} {\bibfnamefont
  {J.}~\bibnamefont {Richard}}, \ and\ \bibinfo {author} {\bibfnamefont
  {E.}~\bibnamefont {Jullo}},\ }\href {\doibase 10.1088/0004-637X/765/1/24}
  {\bibfield  {journal} {\bibinfo  {journal} {Astrophys. J.}\ }\textbf
  {\bibinfo {volume} {765}},\ \bibinfo {pages} {24} (\bibinfo {year} {2013})},\
  \Eprint {http://arxiv.org/abs/1209.1391} {arXiv:1209.1391 [astro-ph.CO]}
  \BibitemShut {NoStop}%
\bibitem [{\citenamefont {Bonnivard}\ \emph {et~al.}(2015)\citenamefont
  {Bonnivard}, \citenamefont {Combet}, \citenamefont {Maurin},\ and\
  \citenamefont {Walker}}]{Bonnivard:2014kza}%
  \BibitemOpen
  \bibfield  {author} {\bibinfo {author} {\bibfnamefont {V.}~\bibnamefont
  {Bonnivard}}, \bibinfo {author} {\bibfnamefont {C.}~\bibnamefont {Combet}},
  \bibinfo {author} {\bibfnamefont {D.}~\bibnamefont {Maurin}}, \ and\ \bibinfo
  {author} {\bibfnamefont {M.~G.}\ \bibnamefont {Walker}},\ }\href {\doibase
  10.1093/mnras/stu2296} {\bibfield  {journal} {\bibinfo  {journal} {Mon. Not.
  Roy. Astron. Soc.}\ }\textbf {\bibinfo {volume} {446}},\ \bibinfo {pages}
  {3002} (\bibinfo {year} {2015})},\ \Eprint {http://arxiv.org/abs/1407.7822}
  {arXiv:1407.7822 [astro-ph.HE]} \BibitemShut {NoStop}%
\end{thebibliography}%

\end{document}